\PassOptionsToPackage{unicode}{hyperref}
\PassOptionsToPackage{hyphens}{url}
\PassOptionsToPackage{dvipsnames,svgnames,x11names}{xcolor}
\documentclass[
]{article}

\usepackage{amsmath,amssymb}
\usepackage{iftex}
\ifPDFTeX
  \usepackage[T1]{fontenc}
  \usepackage[utf8]{inputenc}
  \usepackage{textcomp} % provide euro and other symbols
\else % if luatex or xetex
  \usepackage{unicode-math}
  \defaultfontfeatures{Scale=MatchLowercase}
  \defaultfontfeatures[\rmfamily]{Ligatures=TeX,Scale=1}
\fi
\usepackage{lmodern}
\ifPDFTeX\else  
\fi
\IfFileExists{upquote.sty}{\usepackage{upquote}}{}
\IfFileExists{microtype.sty}{% use microtype if available
  \usepackage[]{microtype}
  \UseMicrotypeSet[protrusion]{basicmath} % disable protrusion for tt fonts
}{}
\makeatletter
\@ifundefined{KOMAClassName}{% if non-KOMA class
  \IfFileExists{parskip.sty}{%
    \usepackage{parskip}
  }{% else
    \setlength{\parindent}{0pt}
    \setlength{\parskip}{6pt plus 2pt minus 1pt}}
}{% if KOMA class
  \KOMAoptions{parskip=half}}
\makeatother
\usepackage{xcolor}
\usepackage[paper=a4paper,left=25mm,right=25mm,top=15mm,bottom=20mm]{geometry}
\makeatletter
\ifx\paragraph\undefined\else
  \let\oldparagraph\paragraph
  \renewcommand{\paragraph}{
    \@ifstar
      \xxxParagraphStar
      \xxxParagraphNoStar
  }
  \newcommand{\xxxParagraphStar}[1]{\oldparagraph*{#1}\mbox{}}
  \newcommand{\xxxParagraphNoStar}[1]{\oldparagraph{#1}\mbox{}}
\fi
\ifx\subparagraph\undefined\else
  \let\oldsubparagraph\subparagraph
  \renewcommand{\subparagraph}{
    \@ifstar
      \xxxSubParagraphStar
      \xxxSubParagraphNoStar
  }
  \newcommand{\xxxSubParagraphStar}[1]{\oldsubparagraph*{#1}\mbox{}}
  \newcommand{\xxxSubParagraphNoStar}[1]{\oldsubparagraph{#1}\mbox{}}
\fi
\makeatother

\providecommand{\tightlist}{%
  \setlength{\itemsep}{0pt}\setlength{\parskip}{0pt}}\usepackage{longtable,booktabs,array}
\usepackage{calc} % for calculating minipage widths
\usepackage{etoolbox}
\makeatletter
\patchcmd\longtable{\par}{\if@noskipsec\mbox{}\fi\par}{}{}
\makeatother
\IfFileExists{footnotehyper.sty}{\usepackage{footnotehyper}}{\usepackage{footnote}}
\makesavenoteenv{longtable}
\usepackage{graphicx}
\makeatletter
\newsavebox\pandoc@box
\newcommand*\pandocbounded[1]{% scales image to fit in text height/width
  \sbox\pandoc@box{#1}%
  \Gscale@div\@tempa{\textheight}{\dimexpr\ht\pandoc@box+\dp\pandoc@box\relax}%
  \Gscale@div\@tempb{\linewidth}{\wd\pandoc@box}%
  \ifdim\@tempb\p@<\@tempa\p@\let\@tempa\@tempb\fi% select the smaller of both
  \ifdim\@tempa\p@<\p@\scalebox{\@tempa}{\usebox\pandoc@box}%
  \else\usebox{\pandoc@box}%
  \fi%
}
\def\fps@figure{htbp}
\makeatother
\NewDocumentCommand\citeproctext{}{}
\NewDocumentCommand\citeproc{mm}{%
  \begingroup\def\citeproctext{#2}\cite{#1}\endgroup}
\makeatletter
 \let\@cite@ofmt\@firstofone
 \def\@biblabel#1{}
 \def\@cite#1#2{{#1\if@tempswa , #2\fi}}
\makeatother
\newlength{\cslhangindent}
\newlength{\csllabelwidth}
\newenvironment{CSLReferences}[2] % #1 hanging-indent, #2 entry-spacing
 {\begin{list}{}{%
  \setlength{\itemindent}{0pt}
  \setlength{\leftmargin}{0pt}
  \setlength{\parsep}{0pt}
  \ifodd #1
   \setlength{\leftmargin}{\cslhangindent}
   \setlength{\itemindent}{-1\cslhangindent}
  \fi
  \setlength{\itemsep}{#2\baselineskip}}}
 {\end{list}}
\usepackage{calc}

\makeatletter
\@ifpackageloaded{tcolorbox}{}{\usepackage[skins,breakable]{tcolorbox}}
\@ifpackageloaded{fontawesome5}{}{\usepackage{fontawesome5}}
\definecolor{quarto-callout-color}{HTML}{909090}
\definecolor{quarto-callout-note-color}{HTML}{0758E5}
\definecolor{quarto-callout-important-color}{HTML}{CC1914}
\definecolor{quarto-callout-warning-color}{HTML}{EB9113}
\definecolor{quarto-callout-tip-color}{HTML}{00A047}
\definecolor{quarto-callout-caution-color}{HTML}{FC5300}
\definecolor{quarto-callout-color-frame}{HTML}{acacac}
\definecolor{quarto-callout-note-color-frame}{HTML}{4582ec}
\definecolor{quarto-callout-important-color-frame}{HTML}{d9534f}
\definecolor{quarto-callout-warning-color-frame}{HTML}{f0ad4e}
\definecolor{quarto-callout-tip-color-frame}{HTML}{02b875}
\definecolor{quarto-callout-caution-color-frame}{HTML}{fd7e14}
\makeatother
\makeatletter
\@ifpackageloaded{caption}{}{\usepackage{caption}}
\AtBeginDocument{%
\ifdefined\contentsname
  \renewcommand*\contentsname{Table of contents}
\else
  \newcommand\contentsname{Table of contents}
\fi
\ifdefined\listfigurename
  \renewcommand*\listfigurename{List of Figures}
\else
  \newcommand\listfigurename{List of Figures}
\fi
\ifdefined\listtablename
  \renewcommand*\listtablename{List of Tables}
\else
  \newcommand\listtablename{List of Tables}
\fi
\ifdefined\figurename
  \renewcommand*\figurename{Figure}
\else
  \newcommand\figurename{Figure}
\fi
\ifdefined\tablename
  \renewcommand*\tablename{Table}
\else
  \newcommand\tablename{Table}
\fi
}
\@ifpackageloaded{float}{}{\usepackage{float}}
\floatstyle{ruled}
\@ifundefined{c@chapter}{\newfloat{codelisting}{h}{lop}}{\newfloat{codelisting}{h}{lop}[chapter]}
\floatname{codelisting}{Listing}

\makeatother
\makeatletter
\@ifpackageloaded{caption}{}{\usepackage{caption}}
\@ifpackageloaded{subcaption}{}{\usepackage{subcaption}}
\makeatother

\usepackage{bookmark}

\IfFileExists{xurl.sty}{\usepackage{xurl}}{} % add URL line breaks if available
\hypersetup{
  pdftitle={Recommendations for visual predictive checks in Bayesian workflow},
  pdfauthor={Teemu Säilynoja; Andrew R. Johnson; Osvaldo A. Martin; Aki Vehtari},
  colorlinks=true,
  linkcolor={blue},
  filecolor={Maroon},
  citecolor={Blue},
  urlcolor={Blue},
  pdfcreator={LaTeX via pandoc}}

\definecolor{quarto-callout-note-color-frame}{HTML}{FFFFFF}

\title{\textbf{Recommendations for visual predictive checks in Bayesian
workflow}}

\author{
  Teemu Säilynoja, Andrew R. Johnson, Osvaldo A. Martin, Aki
Vehtari,  \\
      \textit{Aalto University, Espoo, Finland}
  }

\begin{document}
\maketitle

\begin{tcolorbox}[enhanced jigsaw, colback=white, opacityback=0, left=2mm, arc=.35mm, bottomrule=.15mm, rightrule=.15mm, breakable, toprule=.15mm, leftrule=.75mm, colframe=quarto-callout-important-color-frame]

This article is intended to be viewed as an HTML, and its latest version
is hosted online
\href{https://teemusailynoja.github.io/visual-predictive-checks}{here}.

\end{tcolorbox}

\begin{tcolorbox}[enhanced jigsaw, colback=white, opacityback=0, left=2mm, arc=.35mm, bottomrule=.15mm, rightrule=.15mm, breakable, toprule=.15mm, leftrule=.75mm, colframe=quarto-callout-important-color-frame]
\begin{minipage}[t]{5.5mm}
\textcolor{quarto-callout-important-color}{\faExclamation}
\end{minipage}%
\begin{minipage}[t]{\textwidth - 5.5mm}

\vspace{-3mm}\textbf{Under Review}\vspace{3mm}

This paper is \href{https://www.journalovi.org/under-review.html}{under
review} on the experimental track of the
\href{https://www.journalovi.org/}{Journal of Visualization and
Interaction}. See the
\href{https://github.com/journalovi/2025-sailynoja-visual-predictive-checks/issues}{reviewing
process}.

\end{minipage}%
\end{tcolorbox}

\begin{tcolorbox}[enhanced jigsaw, colback=white, opacityback=0, left=2mm, arc=.35mm, bottomrule=.15mm, rightrule=.15mm, breakable, toprule=.15mm, leftrule=.75mm, colframe=quarto-callout-note-color-frame]

\vspace{-3mm}\textbf{Abstract}\vspace{3mm}

\subparagraph{Introduction}\label{introduction}

A key step in the Bayesian workflow for model building is the graphical
assessment of model predictions, whether these are drawn from the prior
or posterior predictive distribution. The goal of these assessments is
to identify whether the model is a reasonable (and ideally accurate)
representation of the domain knowledge and/or observed data. There are
many commonly used visual predictive checks which can be misleading if
their implicit assumptions do not match the reality. Thus, there is a
need for more guidance for selecting, interpreting, and diagnosing
appropriate visualizations. As a visual predictive check itself can be
viewed as a model fit to data, assessing when this model fails to
represent the data is important for drawing well-informed conclusions.

\subparagraph{Demonstration}\label{demonstration}

We present recommendations for appropriate visual predictive checks for
observations that are: continuous, discrete, or a mixture of the two. We
also discuss diagnostics to aid in the selection of visual methods.
Specifically, in the detection of an incorrect assumption of
continuously-distributed data: identifying when data is likely to be
discrete or contain discrete components, detecting and estimating
possible bounds in data, and a diagnostic of the goodness-of-fit to data
for density plots made through kernel density estimates.

\subparagraph{Conclusion}\label{conclusion}

We offer recommendations and diagnostic tools to mitigate ad-hoc
decision-making in visual predictive checks. These contributions aim to
improve the robustness and interpretability of Bayesian model criticism
practices.

\end{tcolorbox}

\begin{tcolorbox}[enhanced jigsaw, colback=white, opacityback=0, left=2mm, arc=.35mm, bottomrule=.15mm, rightrule=.15mm, breakable, toprule=.15mm, leftrule=.75mm, colframe=quarto-callout-note-color-frame]

\vspace{-3mm}\textbf{Materials, Authorship, License, Conflicts}\vspace{3mm}

\subparagraph{Research materials}\label{research-materials}

Source files of this article, as well as supplementary materials
providing additional detail into the examples and case studies shown in
this article, can be found at
\url{https://teemusailynoja.github.io/visual-predictive-checks}.

\subparagraph{Authorship}\label{authorship}

\begin{itemize}
\tightlist
\item
  \textbf{Teemu Säilynoja}: Conceptualization, Methodology, Software,
  Investigation, Writing - Review \& Editing, Visualization
\item
  \textbf{Andrew R. Johnson}: Conceptualization, Writing - Review \&
  Editing
\item
  \textbf{Osvaldo A. Martin}: Writing - Review \& Editing, Visualization
\item
  \textbf{Aki Vehtari}: Conceptualization, Methodology, Writing - Review
  \& Editing, Supervision, Funding acquisition.
\end{itemize}

\subparagraph{License}\label{license}

This work is licensed under a
\href{http://creativecommons.org/licenses/by/4.0/}{Creative Commons
Attribution 4.0 International License}.

\subparagraph{Conflicts of interest}\label{conflicts-of-interest}

The authors declare that there are no competing interests.

\end{tcolorbox}

\section{Introduction}\label{introduction-1}

Assessing the sensibility of model predictions and their fit to
observations is a key part of most model building workflows. These
assessments may reveal that the model predictions poorly represent (or
replicate) the observed data, prompting the modeller to either improve
their model, or adjust their confidence in the predictions accordingly.
In this paper, we focus on examples rising from Bayesian workflows
(\citeproc{ref-gabry_visualization_2019}{Gabry et al. 2019};
\citeproc{ref-gelman_bayesian_2020}{Gelman et al. 2020}), such as
posterior predictive checking {[}PPC; Box
(\citeproc{ref-box_sampling_1980}{1980});Rubin
(\citeproc{ref-rubin_bayesianly_1984}{1984}); Gelman, Xiao-Li, and Stern
(\citeproc{ref-gelman_posterior_1996}{1996}); Gelman et al.
(\citeproc{ref-gelman_bayesian_2013}{2013}){]}, and visualizations of
data, posterior, and posterior predictive distributions. Visualization
of data distribution is useful even if we do not do any modeling. In
Bayesian inference, the posterior distribution presents the uncertainty
in the parameter values after conditioning on data, and the posterior
predictive distribution presents the uncertainty in the predictions for
new observations. The posterior inference is often performed using Monte
Carlo methods (\citeproc{ref-strumbelj_past_2024}{Štrumbelj et al.
2024}), and posterior and posterior predictive distributions are then
represented with draws from these distributions. In the same way as we
visualize data distributions, we can visualize posterior and posterior
predictive distributions. In posterior predictive checking, we compare
the data distribution to the posterior predictive distribution. If these
distributions are not similar, the model is misspecified and we should
consider improving the model. These visualizations are also applicable
to prior predictive checking, where one might compare prior predictive
samples to some reference values rising from domain knowledge
(\citeproc{ref-wesnerChoosingPriorsBayesian2021}{Wesner and Pomeranz
2021}), and to cross-validation predictive checking where data are
compared to cross-validation predictive distributions to avoid double
use of data.

The purpose of this paper is to illustrate common issues in visual
predictive checking, to provide better recommendations for different use
cases, and to provide diagnostics to automatically warn if useless or
misleading visualization is used. Together we have decades of experience
in teaching visual predictive checking and helping modellers using
popular visual predictive checking tools such as \texttt{bayesplot}
(Gabry and Mahr (\citeproc{ref-gabry_plotting_2022}{2024a})) and
\texttt{ArviZ} (Kumar et al. (\citeproc{ref-kumar_arviz_2019}{2019})).
We have seen many times where students and modellers use commonly
recommended visualizations without realizing that in their context, the
specific visualization is the wrong one and either useless or even
misleading. Backed up with arguments from recent studies in uncertainty
visualization, we provide a summary of methods we recommend. These
methods are aimed to offer an informed basis for decision-making, be it
for the modeller to improve on the model or for the end user to set
expectations on the performance of the model.

The visualizations we discuss should be considered as broadly-applicable
tools aimed to give insight into the overall goodness of model fit to
the observations, as well as to possible deviations in certain parts of
the predictive density. Most modeling workflows benefit from additional
visualizations tailored to their specific use cases, as there is no
general visual or numeric check that would reveal every aspect of the
predictive capabilities of any given model. Some commonly-used methods
for visualizing model predictions have risen from comparisons of
continuous variables, and as such have varying degrees of usability for
assessing predictions with discrete values. Some may not work with all
continuous cases, in case of sharp upper or lower bounds in the set of
possible values the prediction can take.

In Section~\ref{sec-continuous-predictive-distributions}, we inspect the
common use of kernel density estimates (KDE) in summarizing the
predictive distribution and the observed data, and show comparisons
between commonly used visualization approaches. We highlight common
cases when these visualizations may hide useful information and propose
an automated goodness-of-fit test to alert the modeller of the presence
of conflict between the data and the visualization. In
Section~\ref{sec-count-data}, we discuss the use of visualizations
assuming continuous data when the data and predictions are in reality
discrete, but have a high number of unique values. We also discuss an
alternate way of visualizing count data, when the number of cases is
large, but still small enough for the modeller to be interested in
attempting to assess the predictions as individual counts. In
Section~\ref{sec-visual-predictive-checks-for-binary-data}, we focus on
workflows involving binary predictions. We showcase tools that expand
the visualizations beyond the typical bar graphs and binned calibration
plots and instead allow for a more robust assessment of the calibration
of the predictive probabilities. In
Section~\ref{sec-visual-predictive-checks-for-categorical-data} and
Section~\ref{sec-visual-predictive-checks-for-ordinal-data}, we show how
the visual predictive checks described for binary predictions can be
extended to discrete predictions with a small or medium number of
individual cases.

\section{Visual predictive checks for continuous
data}\label{sec-continuous-predictive-distributions}

\begin{figure}

\begin{minipage}{0.33\linewidth}

\centering{

\pandocbounded{\includegraphics[keepaspectratio]{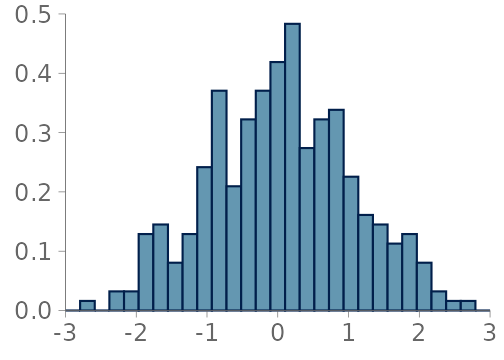}}

}

\subcaption{\label{fig-intro-triplet-histogram}Histogram}

\end{minipage}%
\begin{minipage}{0.33\linewidth}

\centering{

\pandocbounded{\includegraphics[keepaspectratio]{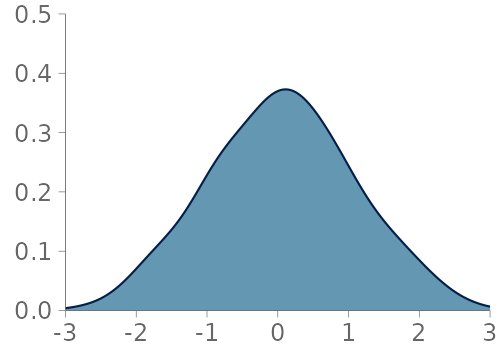}}

}

\subcaption{\label{fig-intro-triplet-kde}KDE Density plot}

\end{minipage}%
\begin{minipage}{0.33\linewidth}

\centering{

\pandocbounded{\includegraphics[keepaspectratio]{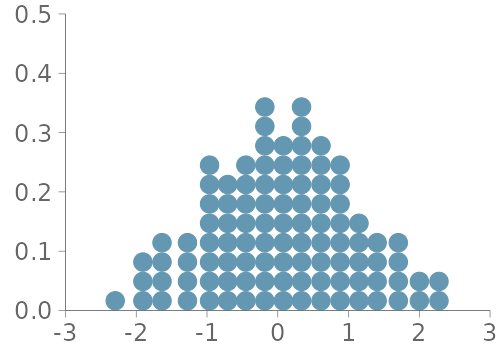}}

}

\subcaption{\label{fig-intro-triplet-qdot}Quantile dot plot}

\end{minipage}%

\caption{\label{fig-intro-triplet}Three visualizations of the same
sample from a smooth continuous distribution. a) Histogram with 30
equally spaced bins. b) Density plot using a Gaussian kernel and
bandwidth \(0.37\) computed through the \textsc{R} implementation of the
method by Sheather and Jones
(\citeproc{ref-sheather_reliable_1991}{1991}). c) Quantile dot plot with
\(100\) quantiles as implemented in \textsc{ggdist} by Kay
(\citeproc{ref-kay_ggdist_2023}{2023}).}

\end{figure}

In this section, we consider visualizations for observations from a
\emph{continuous} or \emph{almost everywhere continuous} distribution.
We focus on three data visualizations; histograms, and KDE-based
continuous density plots, as they are the two most commonly used
visualizations for summarizing unidimensional distributions, and
quantile dot plots (\citeproc{ref-kay_when_2016}{Kay et al. 2016}) as we
see that they offer a useful alternative that we recommend in many
cases. Table~\ref{tbl-pros-cons} summarizes the main advantages and
disadvantages of these three visualizations, and
Figure~\ref{fig-intro-triplet} shows a side-by-side comparison of these
visualizations. Histograms and density plots are implemented in various
commonly used software packages for data visualization and are the two
most common choices for initial visual PPCs (Wickham
(\citeproc{ref-wickham_ggplot2_2016}{2016}); Gabry and Mahr
(\citeproc{ref-gabry_plotting_2022}{2024a}); Kumar et al.
(\citeproc{ref-kumar_arviz_2019}{2019}); Kay
(\citeproc{ref-kay_ggdist_2023}{2023})).

\begin{longtable}[]{@{}
  >{\raggedright\arraybackslash}p{(\linewidth - 4\tabcolsep) * \real{0.1086}}
  >{\raggedright\arraybackslash}p{(\linewidth - 4\tabcolsep) * \real{0.3122}}
  >{\raggedright\arraybackslash}p{(\linewidth - 4\tabcolsep) * \real{0.5747}}@{}}
\toprule\noalign{}
\begin{minipage}[b]{\linewidth}\raggedright
Visualization
\end{minipage} & \begin{minipage}[b]{\linewidth}\raggedright
Advantages
\end{minipage} & \begin{minipage}[b]{\linewidth}\raggedright
Disadvantages
\end{minipage} \\
\midrule\noalign{}
\endfirsthead
\toprule\noalign{}
\begin{minipage}[b]{\linewidth}\raggedright
Visualization
\end{minipage} & \begin{minipage}[b]{\linewidth}\raggedright
Advantages
\end{minipage} & \begin{minipage}[b]{\linewidth}\raggedright
Disadvantages
\end{minipage} \\
\midrule\noalign{}
\endhead
\bottomrule\noalign{}
\tabularnewline
\caption{Summary of advantages and disadvantages of the three
distribution visualizations discussed in the
article.}\label{tbl-pros-cons}\tabularnewline
\endlastfoot
Histogram & \begin{minipage}[t]{\linewidth}\raggedright
\begin{itemize}
\tightlist
\item
  Familiar to most users.
\item
  Can show discontinuities with right binning choice for the data.
\end{itemize}
\end{minipage} & \begin{minipage}[t]{\linewidth}\raggedright
\begin{itemize}
\tightlist
\item
  Requires users to select bin width.
\item
  Challenges in comparing multiple histograms.
\item
  Artefacts with non-discrete valued distributions.
\end{itemize}
\end{minipage} \\
KDE plot & \begin{minipage}[t]{\linewidth}\raggedright
\begin{itemize}
\tightlist
\item
  Familiar to most users.
\item
  Clear visual summary when sample adequately smooth.
\item
  Easy to stack multiple plots for comparison.
\end{itemize}
\end{minipage} & \begin{minipage}[t]{\linewidth}\raggedright
\begin{itemize}
\tightlist
\item
  Tendency to over smooth.
\item
  Challenges with bounded data and discontinuities.
\item
  Selected kernel bandwidth can greatly impact the visualization.
\end{itemize}
\end{minipage} \\
Quantile dot plot & \begin{minipage}[t]{\linewidth}\raggedright
\begin{itemize}
\tightlist
\item
  Dot position adaptive to local smoothness of sample.
\item
  Easily quantifiable tail probability estimation.
\end{itemize}
\end{minipage} & \begin{minipage}[t]{\linewidth}\raggedright
\begin{itemize}
\tightlist
\item
  Not familiar to most users.
\item
  If area fixed and dots are restricted to circles, the aspect ratio is
  locked, unless vertical space is added between dots.
\item
  Exceptions for discrete distributions need to be implemented.
\end{itemize}
\end{minipage} \\
\end{longtable}

\subsubsection{Histogram}\label{histogram}

Here, we consider the most commonly used histograms, which are centered
over the data, and where all of the bins have an equal width (see an
example in Figure~\ref{fig-intro-triplet-histogram}). These bin widths
are either chosen by the modeller, or through a heuristic, such as the
Freedman-Diaconis rule (\citeproc{ref-freedman_histogram_1981}{Freedman
and Diaconis 1981}). Often, instead of choosing the bin width directly,
the modeller inputs a desired amount of bins and an equal partitioning
of the data range is computed with an adequate number of breaks.

\subsubsection{KDE-based density plot}\label{kde-based-density-plot}

The most typical KDE-based density plot, often simply called a density
plot, is a line or filled area representing a density approximation
obtained through convolution of the data with some kernel function (see
an example in Figure~\ref{fig-intro-triplet-kde}). Most commonly, a
Gaussian kernel is used for the approximation, and a kernel bandwidth is
selected through one of the widely used algorithms
(\citeproc{ref-sheather_reliable_1991}{Sheather and Jones 1991};
\citeproc{ref-scott_multivariate_1992}{Scott 1992};
\citeproc{ref-silverman_density_1986}{Silverman 2018}). As the default
implementation of the KDE plot in most visualization software packages
is usually enough to produce aesthetically pleasing smooth summaries of
the data, KDE plots are a very popular method of visually summarizing
data.

\subsubsection{Quantile dot plot}\label{quantile-dot-plot}

The quantile dot plot (\citeproc{ref-kay_when_2016}{Kay et al. 2016}),
is a dot plot, where a set number of the observed quantiles---we use one
hundred---are visualized instead of the observations themselves (see an
example in Figure~\ref{fig-intro-triplet-qdot}). This results in a
visual density estimator with a low bias and high variance similar to
that of a histogram (\citeproc{ref-wilkinson_dot_1999}{Wilkinson 1999}).
Kay et al. (\citeproc{ref-kay_when_2016}{2016}) show that a quantile dot
plot with one hundred quantiles performs very similarly to a KDE-based
density plot for tasks of estimating probabilistic predictions from
visualizations. Compared to KDEs, quantile dot plots have the added
benefit of allowing for fast visual probability estimation in the tails
of the distribution. In our experience, and as shown in
Section~\ref{sec-point-mass} and
Section~\ref{sec-density-function-with-steps} in case of discontinuities
and outliers, quantile dot plots also offer a better visualization than
kernel density plots and histograms.

\subsubsection{Advantages and disadvantages of the three
visualizations}\label{advantages-and-disadvantages-of-the-three-visualizations}

In Table~\ref{tbl-pros-cons} we have collected what we see as the main
advantages and disadvantages of histograms, KDE density plots, and
quantile dot plots for visualizing observations and assessing the
quality of model predictions. Both the histogram and the KDE density
plot have commonly recognized drawbacks, summarized by a trade-off
between over and under smoothing.

The choice of bin width and breakpoints causes histograms to suffer from
binning artefacts (\citeproc{ref-dimitriadis_stable_2021}{Dimitriadis,
Gneiting, and Jordan 2021}). Additionally, comparing the histogram of
observations to multiple histograms visualizing predictive samples may
be difficult.

Density plots from KDEs, as shown in
Section~\ref{sec-density-function-with-steps},
Section~\ref{sec-bounded-density}, and Section~\ref{sec-point-mass},
have a tendency to either hide details such as discontinuities in the
observation distribution if the chosen bandwidth is too large, or
over-fit to the sample when the bandwidth is too small. When
over-fitting, comparing multiple density plots may be difficult, as
there is more variation on the estimate of the underlying distribution.
These drawbacks are typically especially common in the standard
implementations of KDE density plots. More specialized software often
implement measures, such as automated boundary detection and less
conservative bandwidth selection algorithms, to mitigate these issues
(\citeproc{ref-kumar_arviz_2019}{Kumar et al. 2019};
\citeproc{ref-kay_ggdist_2023}{Kay 2023}).

Based on our experience, we prefer quantile dot plots in many cases, as
they offer a low overhead solution to plotting the data without defining
a suitable binning and are flexible enough to represent distributions
with long tails.

\subsection{Assessing the goodness-of-fit of a density
visualization}\label{sec-assessing-the-goodness-of-fit}

As KDE plots, histograms, and quantile dot plots all produce a
visualization of a density, we can assess the representativeness of the
visualization by assessing the goodness-of-fit of the implied density
estimate to the data being visualized. To assess the goodness-of-fit of
a density estimate \(\hat f\), we test for the uniformity of the
probability integral transformed (PIT) data \(\lbrace
x_1,\dots, x_N\rbrace\) when the density approximator is used for the
transform. The PIT value, \(u_i\), of \(x_i\) w.r.t. a density estimator
\(\hat f\) is defined as the cumulative distribution value,
\begin{equation}\phantomsection\label{eq-pit-definition}{
u_i = \int_{-\infty}^{x_i}\hat f(x)\,dx.
}\end{equation}

The underlying principle of this goodness-of-fit test is that, when
computed with regard to the true underlying density \(f\) s.t.
\(x_i \sim f\), the PIT values would satisfy
\(u_n \sim \mathbb U(0,1)\), where \(\mathbb U(0,1)\) is the standard
uniform distribution.

Our uniformity test of choice, due to its graphical representation that
further enhances the explainability of the results, is the graphical
uniformity test proposed by Säilynoja, Bürkner, and Vehtari
(\citeproc{ref-sailynoja_graphical_2022}{2022}). This test provides
simultaneous \(1 - \alpha\) level confidence bands for the uniformity of
the empirical cumulative distribution function (ECDF) of the PIT values.
In the graphical test, the PIT-ECDF and the simultaneous confidence
intervals are evaluated at a set of equidistant points along the unit
interval. An example of these simultaneous confidence intervals for the
ECDF of the PIT values, \(F_{\text{PIT}}\) is provided in
Figure~\ref{fig-kde-continuous-unbounded-ecdf}. Alternatively,
Figure~\ref{fig-kde-continuous-unbounded-ecdf-diff} shows the ECDF
difference version of the same information, that is, the vertical axis
of the plot is scaled to show the difference of the observed ECDF to the
theoretical expectation, \(F_{\text{PIT}}(x) - x\). Especially for large
samples, this transformation allows for an easier assessment of the
values near the ends of the unit interval.

As goodness-of-fit testing does not greatly increase the computational
cost of constructing the visualizations, we propose implementing
automated testing and recommendations for users to consider alternative
visualization techniques when the fit of the chosen visualization isn't
satisfactory.

Next, we introduce how PIT can be implemented for the three
visualizations.

\subsubsection{PIT for KDE density
plots}\label{pit-for-kde-density-plots}

In theory, a KDE defines a proper probability distribution, and the PIT
could be directly computed as in Equation~\ref{eq-pit-definition}.
However, in practice the KDE density visualizations are often showing a
truncated version of the KDE, limited to an interval centered around the
domain of the observed data. For PIT computation, we limit the
integration to the displayed range and normalize the results to make the
truncated KDE integrate to one.

Most software implementations evaluate the KDE on a dense grid spanning
the domain of the observed data, thus allowing numerical PIT computation
by extracting the density values on these evaluation points and carrying
out the aforementioned normalization to obtain PIT values.

\subsubsection{PIT for histograms}\label{pit-for-histograms}

For a histogram with equal bins of width \(h\), we define the PIT as

\begin{align}
    \text{PIT}(x) = h\sum_{j=1}^{J} f_j + (x - l_{J+1})f_{J+1},
\end{align}

where \(J = \max_j \lbrace r_j \leq x\rbrace\), and \(l_j\) and \(r_j\)
are the left and right ends of the \(j\)th bin.

Again, the relative densities and boundaries of these bins are usually
readily available in the software used for constructing these
visualizations, and thus implementing the transform for histograms is a
relatively simple task.

\subsubsection{PIT for quantile dot
plots}\label{pit-for-quantile-dot-plots}

As the quantile dot plots are discrete, the PIT values computed
according to Equation~\ref{eq-pit-definition} are not guaranteed to be
uniformly distributed. As demonstrated by Früiiwirth-Schnatter
(\citeproc{ref-fruiiwirth-schnatter_recursive_1996}{1996}), for a
discrete random variable \(X\), one can employ a randomized
interpolation method to obtain uniformly distributed values,

\begin{align}
  u(x) = \alpha P(X \leq x) + (1-\alpha) P(X \leq (x - 1)),
\end{align} where \(\alpha \sim \mathbb U(0,1)\).

For quantile dot plots, where \(n_q\) is the number of quantiles,
\(c_k\) the horizontal position of the center of the \(k\)th dot and
\(r\) the radius of the dots, we define a randomized PIT as

\begin{align}
    \text{PIT}(x) \sim \mathbb U\left(\frac{l(x)}{n_q}, \frac{u(x)}{n_q}\right) ,
\end{align}§

where \(n_q\) is the amount of quantiles, that is, dots, in the plot,
and \(l(x)\) and \(u(x)\) are the quantile indices satisfying

\begin{align}
    u(x) = \min\lbrace k \in \{1,\dots,n_q\} \mid x < c_k - r\rbrace,
\end{align}

and

\begin{align}
    l(x) = \begin{cases}
        0,&\quad\text{if }u(x) = 1,\\
        \min\lbrace k \in \{1,\dots,n_q\} \mid |c_k - x| \leq
        r\rbrace,&\quad\text{if }\exists \, k \text{ s.t. } |c_k - x| \leq r,\\
        \max\lbrace k \in \{1,\dots,n_q\} \mid x \geq c_k - r\rbrace,&\quad\text{otherwise}.
    \end{cases}
\end{align}

That is, we consider the left and right edges of the dots, \(c_k - r\),
and \(c_k + r\) respectively. Now the PIT value of \(x\) is limited from
above by the smallest quantile dot fully to the right of \(x\). The
lower limit of the PIT value is determined in three cases. First, for
those \(x\) that are to the left of all of the quantile dots, the lower
limit of the PIT value is zero. Second, if \(x\) lies between the left
and right edges of one or more quantile dots, the lower limit is
determined by the smallest of these quantiles. Finally, when \(x\) is
not between the edges of any single quantile dot, the largest dot fully
to the left of \(x\) determines the lower limit of the PIT value.

Again, as the centre points and radius of the dots are required for
constructing the visualizations in the first place, implementing the PIT
computations is relatively straightforward.

\subsection{Continuous valued
observations}\label{sec-almost-everywhere-continuous}

When the observation density is smooth and unbounded, and the
practitioner uses visualizations aimed at continuous distributions,
misrepresenting the data is less common. The main benefit of
goodness-of-fit testing for visualizations arises, when the assumptions
in the visualization do not meet the properties of the underlying
observation distribution. Sometimes data, that may seem continuous at
first glance, proves problematic for KDEs to visualize.

\begin{figure}

\begin{minipage}{0.33\linewidth}

\centering{

\pandocbounded{\includegraphics[keepaspectratio]{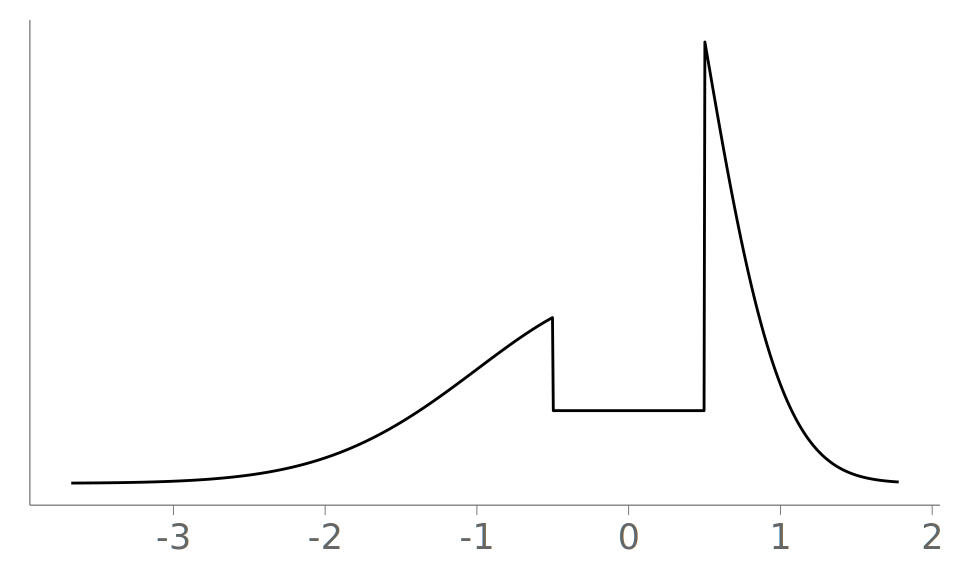}}

}

\subcaption{\label{fig-true-step}Density with steps}

\end{minipage}%
\begin{minipage}{0.33\linewidth}

\centering{

\pandocbounded{\includegraphics[keepaspectratio]{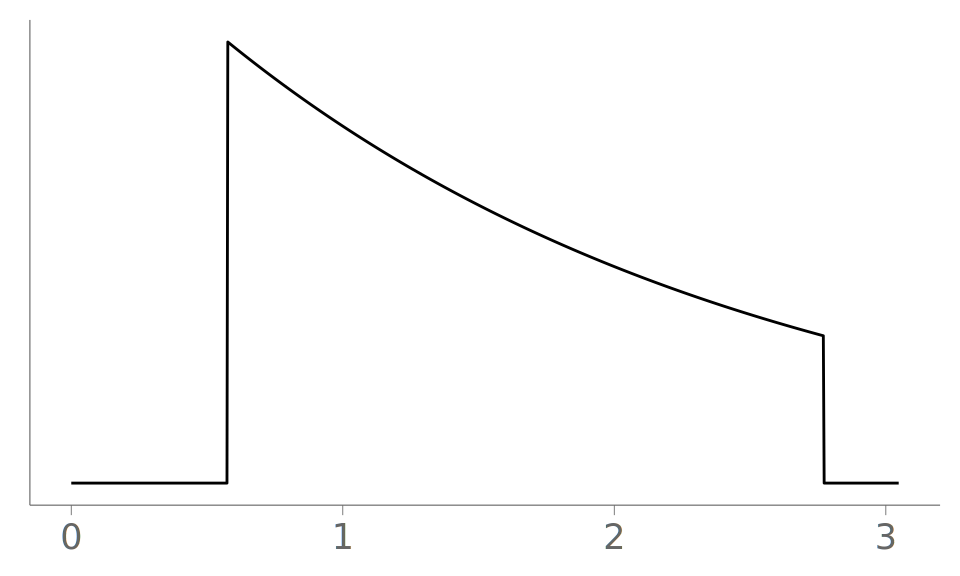}}

}

\subcaption{\label{fig-true-bounded}Strictly bounded density}

\end{minipage}%
\begin{minipage}{0.33\linewidth}

\centering{

\pandocbounded{\includegraphics[keepaspectratio]{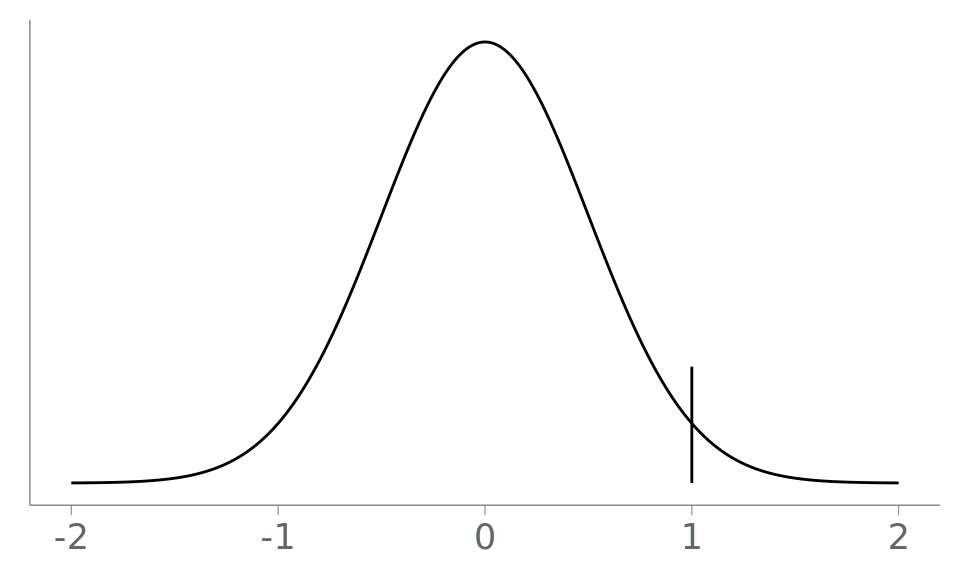}}

}

\subcaption{\label{fig-true-point-mass}Density with a point mass}

\end{minipage}%

\caption{\label{fig-true-densities}True densities for the examples used
in Section~\ref{sec-density-function-with-steps},
Section~\ref{sec-bounded-density}, and Section~\ref{sec-point-mass}.
Each of these examples poses a challenge for KDE density plots and
histograms, and avoiding misrepresenting the data requires special
attention.}

\end{figure}

Below, we will first look at the ideal case of a continuously valued
observation with a smooth and unbounded underlying distribution, and
then step through three examples of continuous valued observations where
the observation distribution is not smooth and visualizing the
observation requires extra attention. The true underlying densities are
visualized in Figure~\ref{fig-true-densities}. In the three later cases,
using the default KDE plot or histogram can hide important details of
the observation, and looking at the issues detected through
goodness-of-fit testing could affect future modeling choices.

\subsection{Smooth and unbounded density}\label{sec-smooth-density}

\begin{figure}

\begin{minipage}{0.33\linewidth}

\centering{

\pandocbounded{\includegraphics[keepaspectratio]{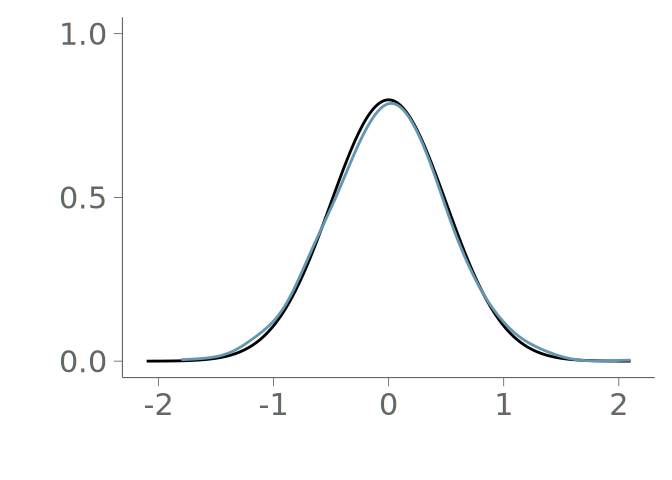}}

}

\subcaption{\label{fig-kde-continuous-unbounded-densityplot}KDE}

\end{minipage}%
\begin{minipage}{0.33\linewidth}

\centering{

\pandocbounded{\includegraphics[keepaspectratio]{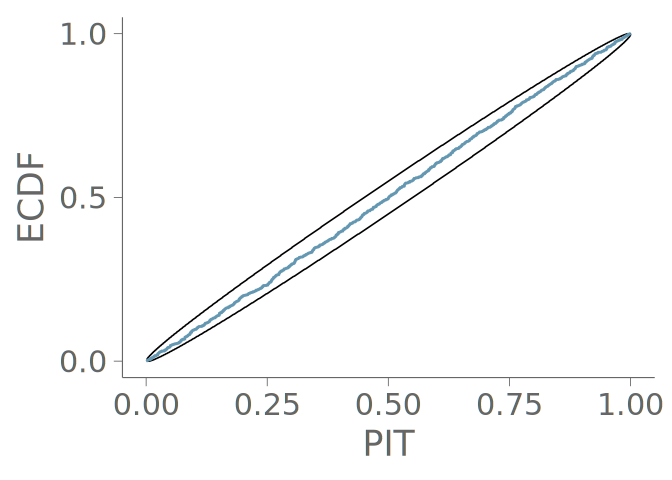}}

}

\subcaption{\label{fig-kde-continuous-unbounded-ecdf}PIT ECDF}

\end{minipage}%
\begin{minipage}{0.33\linewidth}

\centering{

\pandocbounded{\includegraphics[keepaspectratio]{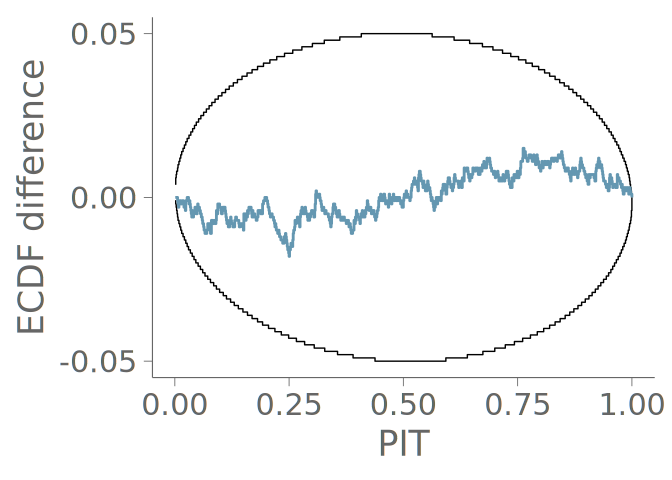}}

}

\subcaption{\label{fig-kde-continuous-unbounded-ecdf-diff}PIT ECDF
difference}

\end{minipage}%

\caption{\label{fig-kde-continuous-unbounded}Visualizing a sample from a
smooth and unbounded density with a kernel density estimate. (a) the KDE
of the sample in blue and the true density in black. (b) The
corresponding PIT ECDF plot with 95\% simultaneous confidence bands for
uniformity. No goodness-of-fit issues are indicated by the plot as the
PIT ECDF stays within the confidence bands. (c) The corresponding PIT
ECDF difference plot, showing the deviation from the expected CDF when
testing against the true distribution. The plot allows for a more
dynamic use of the plotting area, making it easier to inspect the
deviations from the expectation of the CDF.}

\end{figure}

\begin{figure}

\begin{minipage}{0.67\linewidth}
\pandocbounded{\includegraphics[keepaspectratio]{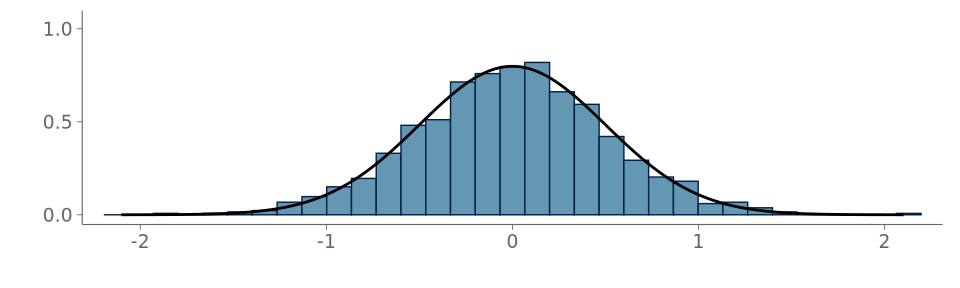}}\end{minipage}%
\begin{minipage}{0.33\linewidth}
\pandocbounded{\includegraphics[keepaspectratio]{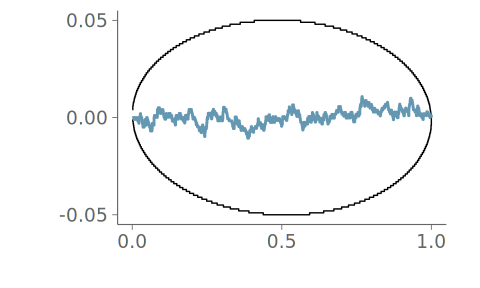}}\end{minipage}%

\caption{\label{fig-histogram-continuous}A histogram of the sample, with
the smooth and unbounded true density overlaid in black. We use the rule
by Freedman and Diaconis (\citeproc{ref-freedman_histogram_1981}{1981})
to determine the number of bins. The PIT ECDF difference plot indicates
a good fit between the visualization and the sample.}

\end{figure}

\begin{figure}

\begin{minipage}{0.67\linewidth}
\pandocbounded{\includegraphics[keepaspectratio]{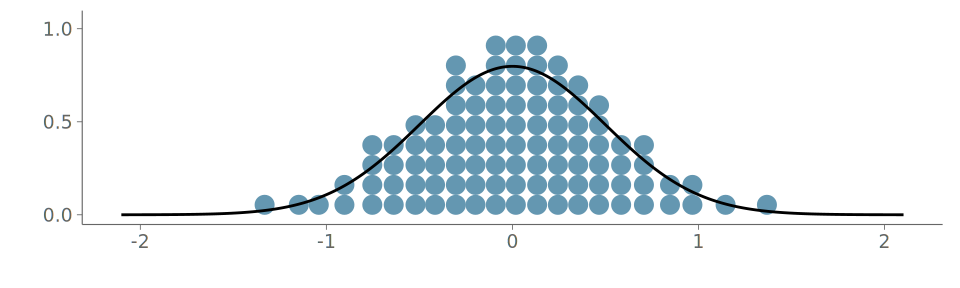}}\end{minipage}%
\begin{minipage}{0.33\linewidth}
\pandocbounded{\includegraphics[keepaspectratio]{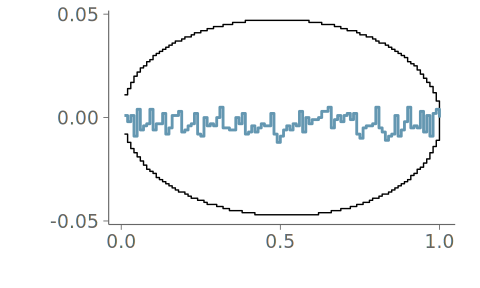}}\end{minipage}%

\caption{\label{fig-qdplot-continuous}The quantile dot plot and the true
density overlaid in black. As with all quantile dot plots in this
article, we show 100 quantiles. The PIT ECDF difference plot indicates a
good fit between the visualization and the sample.}

\end{figure}

When the underlying distribution is smooth and unbounded, large issues
in goodness-of-fit of the discussed three visualizations are rare.
Figure~\ref{fig-kde-continuous-unbounded-densityplot} includes the KDE
plot for an observation of \(1000\) standard normally distributed
values. Figures Figure~\ref{fig-kde-continuous-unbounded-ecdf} and
Figure~\ref{fig-kde-continuous-unbounded-ecdf-diff} show the
corresponding goodness-of-fit test with the PIT ECDF values well within
the 95\% simultaneous confidence intervals for uniformity.
Figure~\ref{fig-histogram-continuous} shows the histogram and the
corresponding goodness-of-fit assessment of the same data. This time, we
show only the ECDF difference version of the graphical test. Again, no
issues are detected. Figure~\ref{fig-qdplot-continuous} in turn shows
the quantile dot plot paired with the corresponding goodness-of-fit
evaluation. Again, no issues are detected. Although one hundred
evaluation points results in a quite smooth visualization, the
discreteness of the PIT ECDF of the quantile dot plot is visible.

\subsection{Density with steps}\label{sec-density-function-with-steps}

When the observation density has steps, that is, points with unequal
one-sided limits, the continuous representation offered by KDEs---that
assume smooth density---may struggle to accurately represent the
discontinuity. In turn, the visualizations provided by histograms and
quantile dot plots are more flexible, but if the location of the step is
not known, an unfortunate histogram binning may introduce large
deviations to the true density values. If the location of the step was
known, a good representation could also be obtained with KDE plots or
histograms by using separate bounded KDEs for the two sides of the step
(see Section~\ref{sec-bounded-density}), or by tailoring the histogram
by aligning the discontinuity to the boundary of two adjacent bins.

To illustrate the challenges of visualizing observations from stepped
densities, we use samples from the following true density \(f\), shown
in Figure~\ref{fig-true-step}:

\begin{align}
f(x) = \begin{cases}
         \frac{2}{5}\Phi(-\frac{1}{2})^{-1}\mathcal{N}(x\mid 0,1), & x \leq -\frac{1}{2}\\
          \frac{1}{5}, & -\frac{1}{2} < x \leq \frac{1}{2}\\
         \frac{2}{5}\Phi(-\frac{1}{4})^{-1}\mathcal{N}(x\mid 0,\frac{1}{4}), & x > \frac{1}{2},
       \end{cases}
\end{align}

this kind of bimodal skewed distribution with low density between the
modes is present for example in the z-scores of published p-values
(\citeproc{ref-van_zwet_significance_2021}{Van Zwet and Cator 2021}).

\begin{figure}

\begin{minipage}{0.67\linewidth}
\pandocbounded{\includegraphics[keepaspectratio]{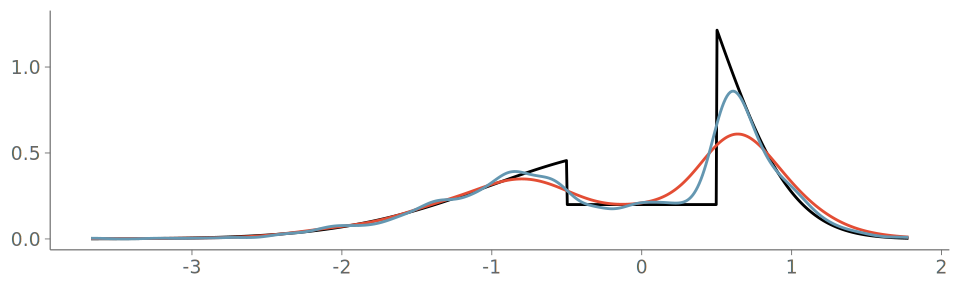}}\end{minipage}%
\begin{minipage}{0.33\linewidth}
\pandocbounded{\includegraphics[keepaspectratio]{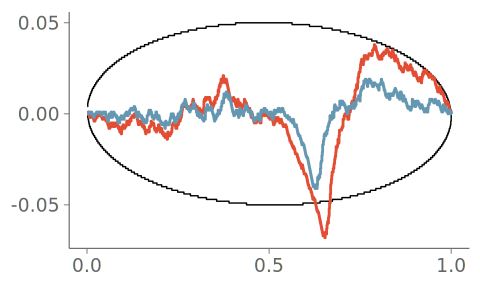}}\end{minipage}%

\caption{\label{fig-kde-step}Two kernel density plots for a continuous
valued sample from a density with steps. In {red}, a KDE with the
bandwidth selected with {Silverman's rule of thumb}
(\citeproc{ref-silverman_density_1986}{Silverman 2018}), and in {blue} a
KDE using the {Sheather-Jones} bandwidth selection method
(\citeproc{ref-sheather_reliable_1991}{Sheather and Jones 1991}), which
results to a smaller bandwidth and a better fit to the sample.}

\end{figure}

Figure~\ref{fig-kde-step} shows the resulting density plot and the
corresponding calibration for two common kernel bandwidth selection
strategies; Silverman's rule of thumb
(\citeproc{ref-silverman_density_1986}{Silverman 2018}), and the
Sheather-Jones (SJ) method
(\citeproc{ref-sheather_reliable_1991}{Sheather and Jones 1991}). Of
these, SJ is expected to give a more robust bandwidth selection to data
from non-Gaussian distributions
(\citeproc{ref-sheather_reliable_1991}{Sheather and Jones 1991}).
Despite this, Silverman's rule of thumb is the default strategy in many
KDE density visualization implementations using Gaussian kernels. As
seen in the figure, both strategies have difficulties representing the
discontinuity in the observation density, and PIT ECDF for the KDE plot
using Silverman's rule of thumb crosses the 95\% simultaneous confidence
interval and is recognized as having significant goodness-of-fit issues.

\begin{figure}

\begin{minipage}{0.67\linewidth}
\pandocbounded{\includegraphics[keepaspectratio]{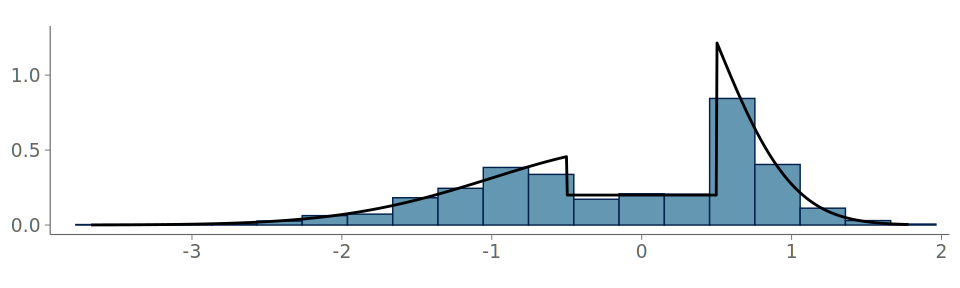}}\end{minipage}%
\begin{minipage}{0.33\linewidth}
\pandocbounded{\includegraphics[keepaspectratio]{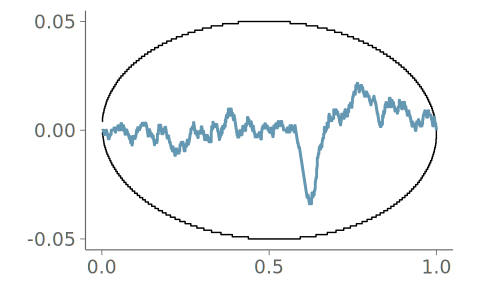}}\end{minipage}%

\caption{\label{fig-histogram-step}Visualizing the continuous valued
sample from a density with steps with a histogram. Here, our bin width
selection algorithm has resulted in an adequate fit to the data,
although the local deviation from the expected CDF is clearly visible in
the PIT ECDF plot.}

\end{figure}

Figure~\ref{fig-histogram-step} shows the visualization and
goodness-of-fit assessment for the same data, when using a histogram.
Although the discontinuity is strongly hinted in the histogram, it is
not located close to a bin boundary and causes significant
goodness-of-fit issues as too much density is placed on the last values
of the low density region preceding the discontinuity.

\begin{figure}

\begin{minipage}{0.67\linewidth}
\pandocbounded{\includegraphics[keepaspectratio]{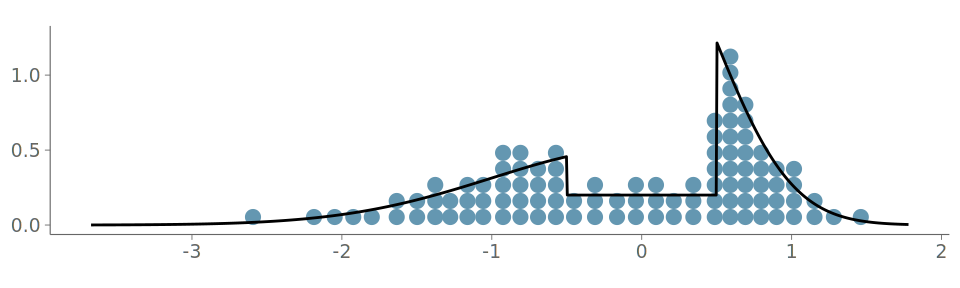}}\end{minipage}%
\begin{minipage}{0.33\linewidth}
\pandocbounded{\includegraphics[keepaspectratio]{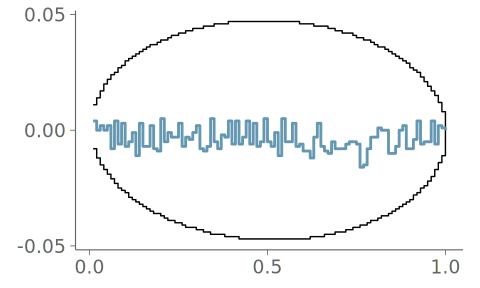}}\end{minipage}%

\caption{\label{fig-qplot-step}Visualizing the continuous valued sample
from a density with steps with a quantile dot plot. The larger step is
clearly visible, and PIT ECDF plot stays within the simultaneous
confidence intervals. The 100 quantiles are able to adapt to the sample
better than the histogram above with a relatively low number of bins
obtained from the used binning algorithm.}

\end{figure}

Lastly, Figure~\ref{fig-qplot-step} shows the same process for a
quantile dot plot. Again, the discontinuities are visible in the plot.
Here, as the visualization follows the ECDF quantiles, the
discontinuities don't cause issues to the goodness-of-fit.

\subsection{Density with strict bounds}\label{sec-bounded-density}

Bounded density functions are very commonplace, and a special case of
the stepped densities discussed above. If the data is known to be
bounded, the knowledge allows for specialized visualizations; for
example, KDE plots with boundary corrections methods, such as boundary
reflection, or limiting the histogram bins to the domain of the density.

A problem arises, when unknowingly visualizing bounded data. Below, we
inspect our three visualizations of interest applied to data from a
truncated exponential distribution with rate parameter \(\lambda = 1\)
(shown in Figure~\ref{fig-true-bounded}). The truncation is to the
central \(80\%\) interval of the untruncated distribution.

\begin{figure}

\begin{minipage}{0.67\linewidth}
\pandocbounded{\includegraphics[keepaspectratio]{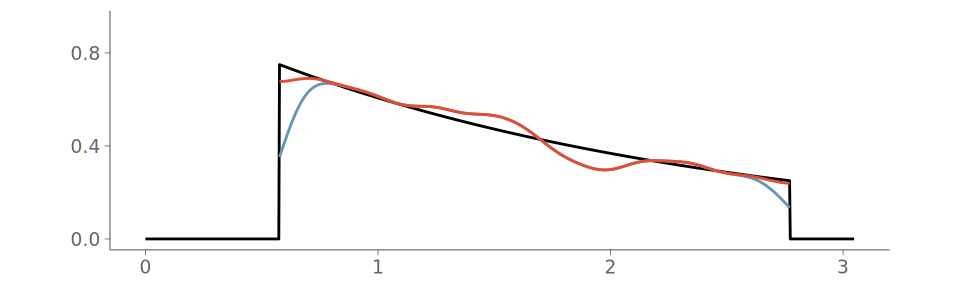}}\end{minipage}%
\begin{minipage}{0.33\linewidth}
\pandocbounded{\includegraphics[keepaspectratio]{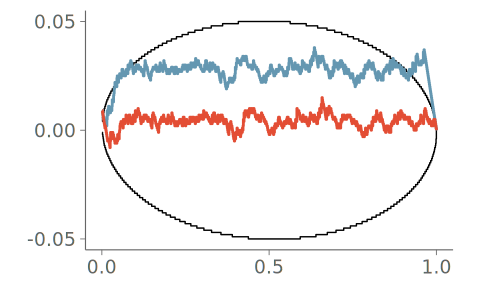}}\end{minipage}%

\caption{\label{fig-kde-bounds}Two kernel density plots for a continuous
valued sample from a bounded density. In {red}, a KDE using a {boundary
correction} and automated boundary detection as implemented in the
ggdist package for R (\citeproc{ref-kay_ggdist_2023}{Kay 2023}). The KDE
in {blue} assumes {unbounded} data and uses no boundary corrections.
Again, the misrepresentation of the sample close to the distribution
boundaries is detected with the graphical goodness-of-fit test.}

\end{figure}

Figure~\ref{fig-kde-bounds} shows a comparison between two KDE plots;
one without any information on the boundedness of the data, the other
made using boundary reflection based on bounds estimated through an
automated boundary detection algorithm implemented in the
\texttt{ggdist} R package (\citeproc{ref-kay_ggdist_2023}{Kay 2023}).
The goodness-of-fit test clearly indicates that the density is
misrepresented near the boundaries by the unbounded KDE plot. In
general, a \(\cap\)-shaped PIT ECDF plot indicates bias towards too
small PIT values, and the strong upward trend at small PIT values
indicates that the estimated density at the left tail is lower than
expected if the data was sampled from the estimated density.

\begin{figure}

\begin{minipage}{0.67\linewidth}
\pandocbounded{\includegraphics[keepaspectratio]{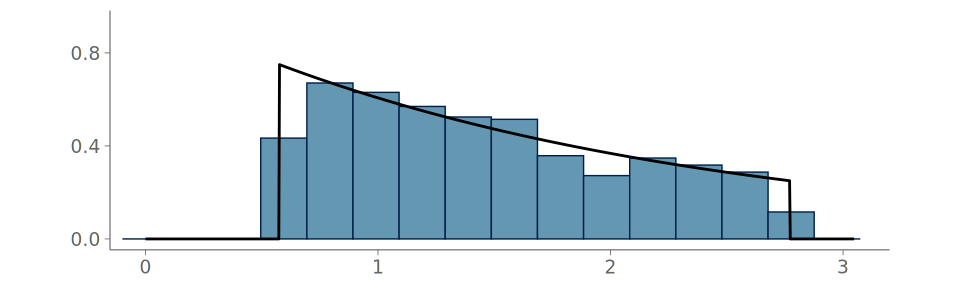}}\end{minipage}%
\begin{minipage}{0.33\linewidth}
\pandocbounded{\includegraphics[keepaspectratio]{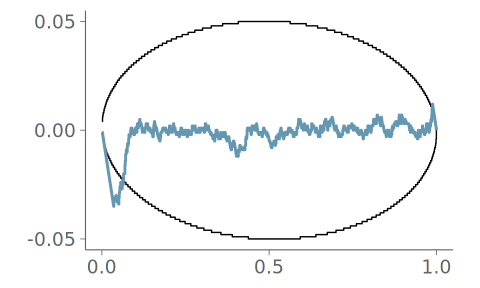}}\end{minipage}%

\caption{\label{fig-histogram-bounds}Visualizing the continuous valued
sample from a bounded density with a histogram. The PIT ECDF line
crosses the simultaneous confidence bands at the extreme PIT values,
indicating issues in representing the boundaries of the data
distribution. The drop in the PIT ECDF close to zero indicates
underestimation of the left bound of the distribution, and respectively,
the smaller upwards peak close to one indicates over estimation of the
right bound.}

\end{figure}

Figure~\ref{fig-histogram-bounds} shows the data visualized with a
histogram without limiting the bins to lie inside the bounds. Again,
bins overlap the discontinuities, causing the goodness-of-fit test to
indicate possible data misrepresentation.

\begin{figure}

\begin{minipage}{0.67\linewidth}
\pandocbounded{\includegraphics[keepaspectratio]{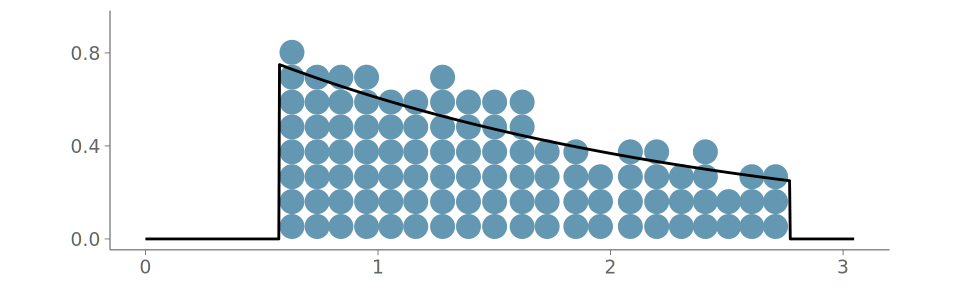}}\end{minipage}%
\begin{minipage}{0.33\linewidth}
\pandocbounded{\includegraphics[keepaspectratio]{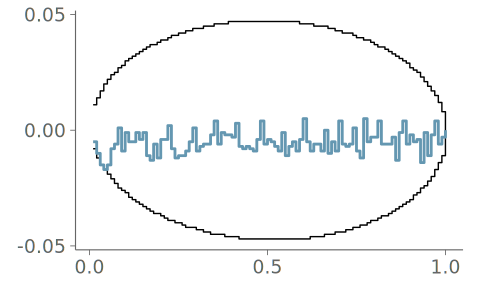}}\end{minipage}%

\caption{\label{fig-qplot-bounds}Visualizing the continuous valued
sample from a bounded density with a quantile dot plot. The quantile
dots are correctly placed within the bounds of the data distribution,
making the quantile dot plot a good alternative for visualizing bounded
data.}

\end{figure}

Figure~\ref{fig-qplot-bounds} shows the data visualized with quantile
dot plot using 100 quantiles. As the quantile dot plot by design isn't
placing any dots outside the data range, the goodness-of-fit is
satisfactory.

\subsection{Density with point masses}\label{sec-point-mass}

The final example of continuous valued observations with discontinuities
in the density is met when one or more point masses are present. These
kinds of observation densities are often met in biomedical and
economical studies, where zero-inflated, but otherwise continuous data
is common (\citeproc{ref-liu_statistical_2019}{Liu et al. 2019}).
Another cause of point masses can be problematic treatment of missing
values, where any missing fields in multidimensional observations are
replaced with some predetermined value, be it zero or some summary
statistic.

Below, we inspect an example following the density depicted in
Figure~\ref{fig-true-point-mass}, where an observed value from the
standard normal distribution is replaced with \(1\) with probability
\(0.2\).

\begin{figure}

\begin{minipage}{0.67\linewidth}
\pandocbounded{\includegraphics[keepaspectratio]{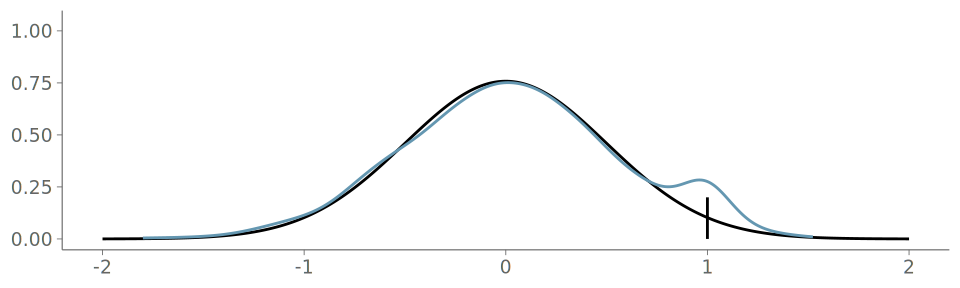}}\end{minipage}%
\begin{minipage}{0.33\linewidth}
\pandocbounded{\includegraphics[keepaspectratio]{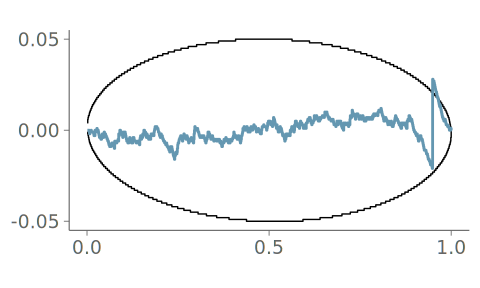}}\end{minipage}%

\caption{\label{fig-kde-point-mass}KDE for a continuous valued sample
from a density with a point mass. The selected bandwidth is too large to
adequately represent the point mass. This is detected by the
goodness-of-fit test.}

\end{figure}

Figure~\ref{fig-kde-point-mass} shows the KDE plot and the corresponding
goodness-of-fit assessment, when the SJ method is used for bandwidth
selection. Although the point mass does show as an additional bump in
the density, the goodness-of-fit test notices that the KDE isn't
flexible enough and alerts us of the underlying point mass with a sharp
jump in the PIT ECDF.

\begin{figure}

\begin{minipage}{0.67\linewidth}
\pandocbounded{\includegraphics[keepaspectratio]{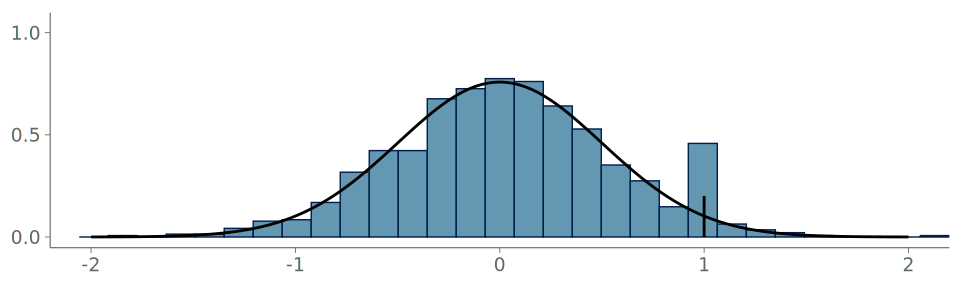}}\end{minipage}%
\begin{minipage}{0.33\linewidth}
\pandocbounded{\includegraphics[keepaspectratio]{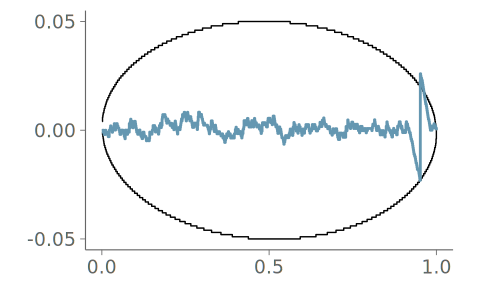}}\end{minipage}%

\caption{\label{fig-histogram-point-mass}Histogram for the continuous
valued sample from a density with a point mass. The selected bin width
is too wide and the misrepresentation of the data is detected by the
goodness-of-fit test.}

\end{figure}

Figure~\ref{fig-histogram-point-mass} shows the same data visualized
using a histogram. Now the discontinuity is arguably more visible in the
visualization, but, as all the bins are of equal width, the PIT ECDF
shows a sharp discontinuity.

\begin{figure}

\begin{minipage}{0.67\linewidth}
\pandocbounded{\includegraphics[keepaspectratio]{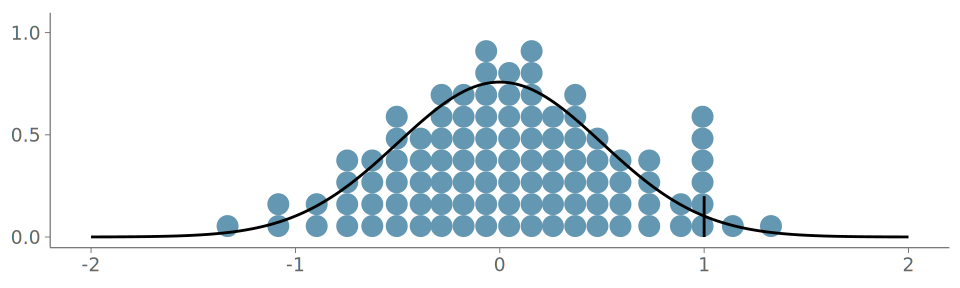}}\end{minipage}%
\begin{minipage}{0.33\linewidth}
\pandocbounded{\includegraphics[keepaspectratio]{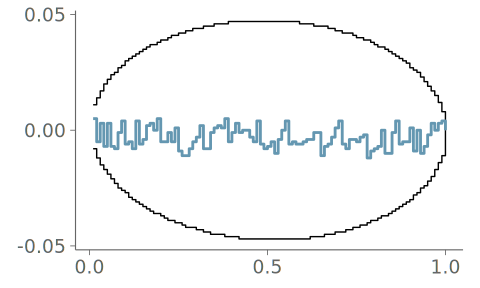}}\end{minipage}%

\caption{\label{fig-qplot-point-mass}Quantile dot plot of the continuous
valued sample from a density with a point mass. The quantiles dots are
able to represent the point mass and the visualization passes the
goodness-of-fit test.}

\end{figure}

Figure~\ref{fig-qplot-point-mass} shows the data visualized with
quantile dot plot using 100 quantiles. Again, the point mass is visible
in the visualization, and the PIT ECDF indicates that the visualization
is representative of the sample.

\subsubsection{Detecting if observation is discrete or mixture of
discrete and
continuous}\label{detecting-if-observation-is-discrete-or-mixture-of-discrete-and-continuous}

In addition to the aforementioned goodness-of-fit test, a fast and
simple to implement method is to simply count unique values in the
observed data. Relatively high counts of repeated values can inform the
practitioner if the observations are discrete or contain point masses,
for example, zero-inflation.

If at least one non-unique value, we move to check for relative
frequencies of repeated values. If the relative frequency of any value
is more than 2\% of the full sample, we consider the possibility that
the data might contain discrete values.

\subsection{Visual predictive checking with overlaid
plots}\label{sec-predictive-checks}

\begin{figure}

\begin{minipage}{0.33\linewidth}

\centering{

\pandocbounded{\includegraphics[keepaspectratio]{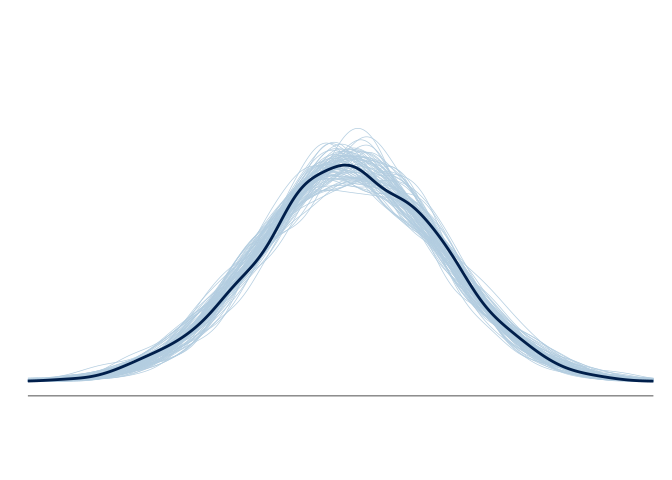}}

}

\subcaption{\label{fig-density-overlay}KDE Overlay}

\end{minipage}%
\begin{minipage}{0.33\linewidth}

\centering{

\pandocbounded{\includegraphics[keepaspectratio]{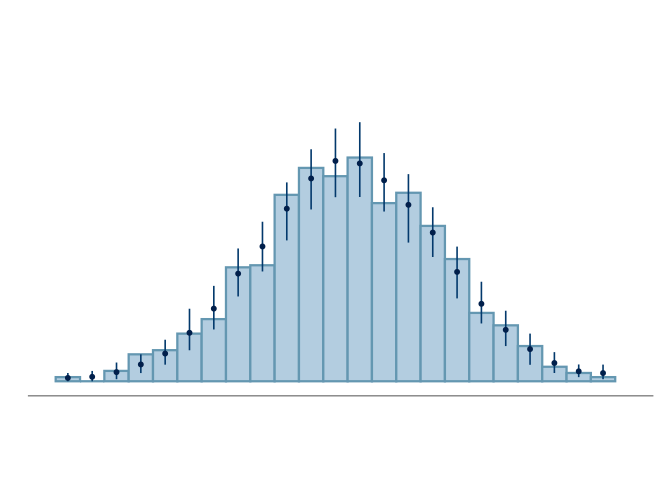}}

}

\subcaption{\label{fig-histogram-overlay}Histogram Overlay}

\end{minipage}%
\begin{minipage}{0.33\linewidth}

\centering{

\pandocbounded{\includegraphics[keepaspectratio]{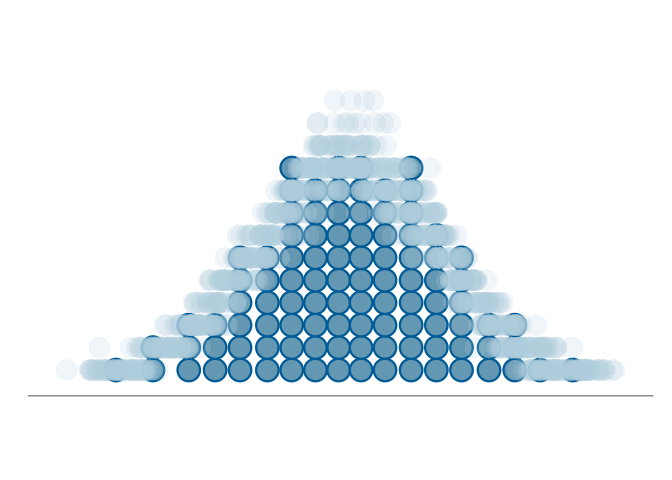}}

}

\subcaption{\label{fig-qdotplot-overlay}Quantile dot plot overlay}

\end{minipage}%

\caption{\label{fig-overlaid-predictive-checks}Overlaid predictive
checks. KDE plots can easily be overlaid and the resulting visualization
is straightforward to read. The histogram for the data is overlaid with
means and central 90\% quantiles of the histogram bins of the posterior
draws, when using the binning from the histogram of the observation. For
the quantile dot plot, the top dot of each stack of dots in each
posterior draw is overlaid to the full quantile dot plot of the
observation, visualizing the variation in the shape of the predictive
distributions.}

\end{figure}

When comparing data and model predictions, one approach is to show
visualization of data and visualization of the draws from the predictive
distribution overlaid in one plot. The most common approach is to plot
overlaid KDE plots, as it is usually easy to distinguish several
overlaid KDE lines from each other. The bandwidth selection plays an
important role in the comparison. On one hand, if the chosen bandwidth
is narrow, the KDEs can show large variation even when repeatedly
drawing from the same distribution. On the other hand, with a wide
bandwidth, the KDE may be too smooth and hide important details. With
histograms, and quantile dot plots, overlaid plots are rarely used, as
the bin boundaries and dot locations depend on the model prediction and
the resulting overlaid figure becomes difficult to read. A common
solution for histograms, shown later in Section~\ref{sec-count-data}, is
to use a shared set of bins for the predictions and data, and to
increase readability by only show the predictions through per bin
summary statistics, such as mean and quantile based intervals. A
drawback with this approach is, that the per bin summaries don't show
dependency between the bins, making it harder to assess the global shape
of the density. To compare multiple quantile dot plots to a reference
plot, we propose overlaying the reference plot with just the top dots of
each stack in the quantile dot plots of the predictive samples.
Figure~\ref{fig-qdotplot-overlay} shows the resulting plot, which
illustrates the variation in the overall form of the predictive
distribution. Again, when summarizing the quantile dot plots of the
predictive draws, we lose information on the dependency between heights
of the stacks and the overall shape of the predictive density.

Overlaid KDE plots are usually easy to interpret---assuming the
smoothness assumptions match the underlying data and draws from the
predictive distribution---although they lack a quantification for when
the differences are significant. For better quantification of the
difference and more robust behavior in the case of non-smooth underlying
distributions, we recommend the use of the graphical PIT-ECDF test
described in Section~\ref{sec-assessing-the-goodness-of-fit}. In
posterior predictive checks, we assess the fit between the observations
and the predictive distribution of the model. The predictive
distribution is not expected to give an exact match, for example the
posterior predictive distribution of a Bayesian model usually has
thicker tails than the true underlying observation distribution. To
avoid using the same observations for both fitting the model and
assessing the predictive performance, we use leave-one-out
cross-validation (LOO-CV). Especially flexible models and small
observations require LOO-CV for more reliable PIT values. If one were to
not use cross-validation, the PIT values would tend to be too close to
\(0.5\) and the PIT ECDF plot would be S-shaped, suggesting many
observations are too close to their respective predictive means. For
fast LOO-CV computation, we recommend using Pareto smoothed importance
sampling (PSIS) (\citeproc{ref-JMLR:v25:19-556}{Vehtari et al. 2024}),
estimates the leave-one-out predictive distributions without the need
for refitting the model for each left out observation. In our
experience, the PIT ECDF plots of both the posterior predictive and
LOO-predictive draws are useful for assessing that the predictive
distribution of the model is well calibrated and the visualizations give
insight to the nature of the possible issues in model fit.

\section{Visual predictive checks for count data}\label{sec-count-data}

In this section, we focus on count data, which although being discrete,
can benefit from both continuous and discrete visualizations. Continuous
visualizations are commonly used for count data and they can sometimes
be a good choice, but there are cases where they can be misleading and
visual predictive checks designed specifically for count data should be
used.

In predictive checking for count data, we focus on two visualizations;
the overlaid KDE plots introduced in
Section~\ref{sec-predictive-checks}, and our version of the rootogram
(\citeproc{ref-kleiber_visualizing_2016}{Kleiber and Zeileis 2016}). In
our experience, the higher the number of distinct counts being
visualized, and the smaller the variance of probabilities of individual
counts is, the more effective KDEs are for visualizing count data,
although one should pay close attention to the effect of the boundedness
on the KDE fit as count data is usually limited to non-negative values
and sometimes also has maximum count limit. Rootograms offer a
visualization which emphasizes the discrete nature of the predictive and
observation distributions. In our experience, rootograms are good for
predictive checking of count data, when the number of distinct counts is
low or there is a sharp changes in the probabilities of consecutive
counts, for example, when the data exhibits zero-inflation (the
probability of \(0\) is much higher than the probability of \(1\)).

In Section~\ref{sec-counts-with-ecdf} we show how, similarly to the
cases in Section~\ref{sec-continuous-predictive-distributions},
goodness-of-fit tests can be used to assess how well the KDE plot
represents the data. In Section~\ref{sec-counts-with-rootogram} we
introduce our version of the rootogram, and demonstrate its use when the
familiar KDE exhibits goodness-of-fit issues. The visual predictive
checks in both Section~\ref{sec-counts-with-ecdf} and
Section~\ref{sec-counts-with-rootogram} follow an example from a
modeling workflow of estimating the effect of integrated pest management
on reducing cockroach levels in urban apartments
(\citeproc{ref-gelman_regression_2020}{Gelman, Hill, and Vehtari 2020}).
We focus on assessing the quality of the predictions of a generalized
linear model with a negative binomial data model for the number of
roaches caught in traps during the original experiment.

\subsection{Visualizing count data with KDE
plots}\label{sec-counts-with-ecdf}

\begin{figure}

\begin{minipage}{0.50\linewidth}
\pandocbounded{\includegraphics[keepaspectratio]{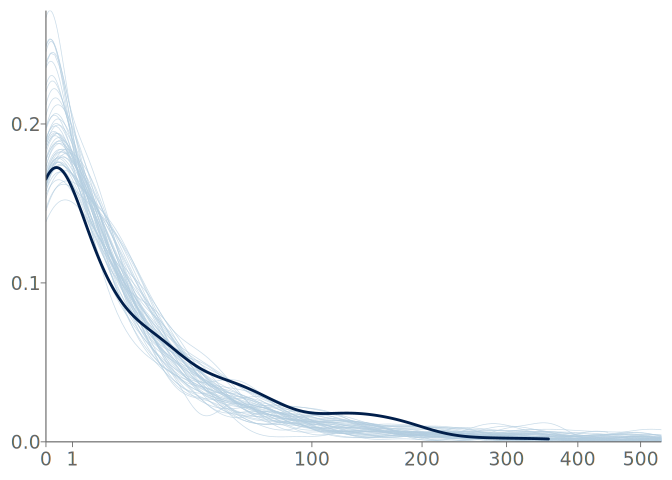}}\end{minipage}%
\begin{minipage}{0.50\linewidth}
\pandocbounded{\includegraphics[keepaspectratio]{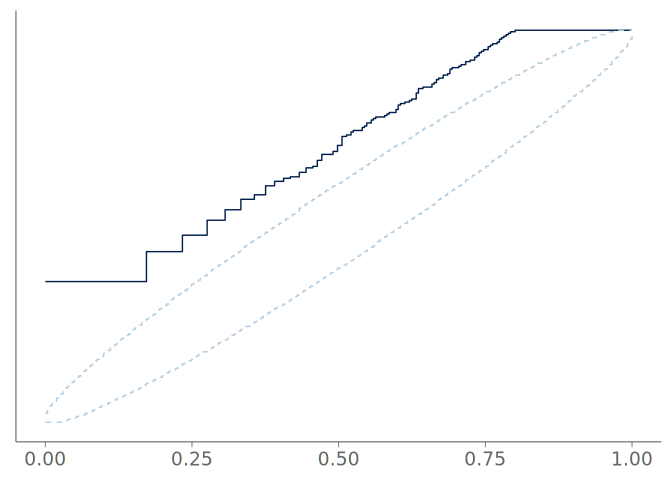}}\end{minipage}%

\caption{\label{fig-high-cardinality-count-data-kde-ecdf}An overlaid KDE
plot of the observed roach counts and the predictive samples. The KDE
density uses the default predictive check implementation of
\texttt{bayesplot}, and is not bounded and has a relatively large kernel
bandwidth. The PIT ECDF shows that the visualization assigns too much
probability mass to small counts, and between the discrete integer
valued observations, resulting in the PIT ECDF staying above the
simultaneous confidence bands.}

\end{figure}

\begin{figure}

\begin{minipage}{0.50\linewidth}
\pandocbounded{\includegraphics[keepaspectratio]{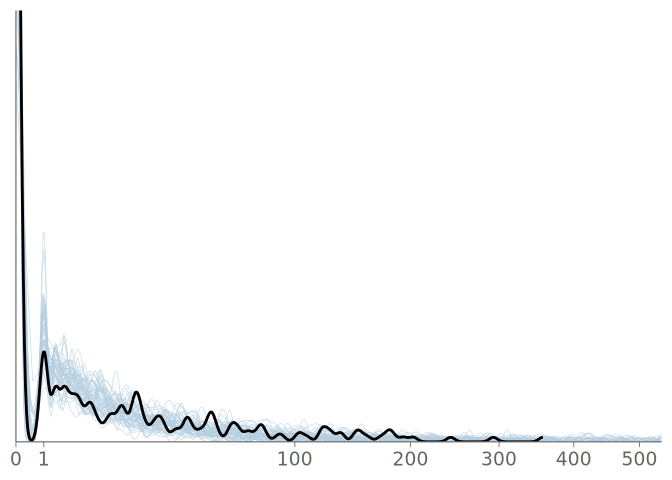}}\end{minipage}%
\begin{minipage}{0.50\linewidth}
\pandocbounded{\includegraphics[keepaspectratio]{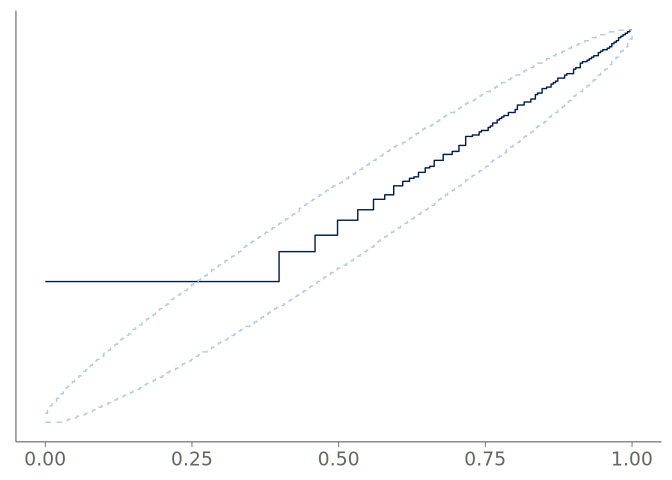}}\end{minipage}%

\caption{\label{fig-small-bandwidth-count-data-kde}An overlaid KDE plot
of the observed roach counts and the predictive samples. The KDE is
bounded at zero and uses the SJ method for bandwidth selection,
resulting in a better fit to the discrete data. Due to zero inflation,
there are numerous ties in the PIT values and the PIT ECDF is outside
the simultaneous confidence bands at zero. We see that for larger
observed counts, the visualization represents the data well.}

\end{figure}

A common approach for count data, especially when the number of unique
discrete values is high, is to assume that the values can be visualized
as if they were from a continuous distribution. This enables the use of
overlaid KDE plots for visual predictive checks.
Figure~\ref{fig-high-cardinality-count-data-kde-ecdf} and
Figure~\ref{fig-small-bandwidth-count-data-kde} illustrate how---after
changing the default bandwidth algorithm---a KDE plot can have a
satisfactory goodness-of-fit to discrete data, and how the PIT ECDF
diagnostic can be used to assess the goodness-of-fit of the continuous
visual representation of the discrete data.

The data in the roach dataset exhibits zero-inflation, and the default
algorithm of the selected visualization software fits an unbounded KDE
with a relatively large bandwidth, resulting in a bad fit shown in
Figure~\ref{fig-high-cardinality-count-data-kde-ecdf}. After using a
left bounded KDE and the SJ bandwidth selection method, we see in
Figure~\ref{fig-small-bandwidth-count-data-kde} that away from zero, the
observation distribution is well summarized by the KDE plot.

A reasonable step in the visual predictive checking for the roaches
model, is to separate the analysis of the long tail and the bulk of the
distribution. One possible approach for analyzing the tail of the
distribution, is to compare how the estimated Pareto shape parameter of
the predictive draws compares to the tail shape of the data. When
comparing the tail shapes, one shouldn't expect an exact fit, as the
predictive distributions typically exhibit thicker tails due to
uncertainty. When moving to analyze the bulk of the distribution, we see
that the number of distinct counts below 90\% quantile is relatively
small, \(48\), and thus we may expect a discrete visual predictive check
to perform well.

\subsection{Rootograms emphasize discreteness of count
data}\label{sec-counts-with-rootogram}

\begin{figure}

\begin{minipage}{0.33\linewidth}

\centering{

\pandocbounded{\includegraphics[keepaspectratio]{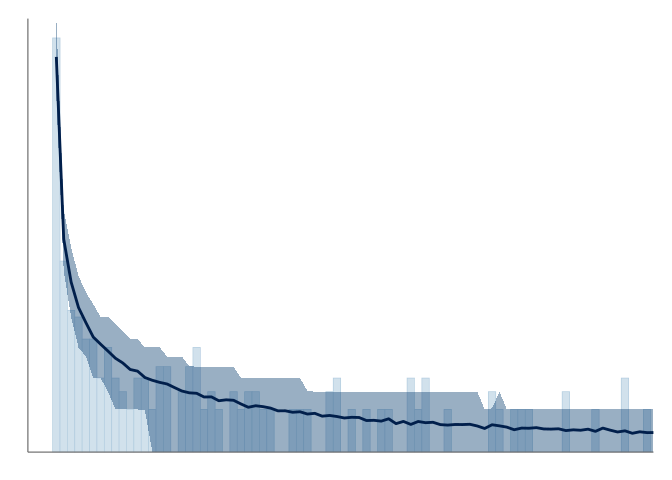}}

}

\subcaption{\label{fig-standing-rootogram}Standing Rootogram}

\end{minipage}%
\begin{minipage}{0.33\linewidth}

\centering{

\pandocbounded{\includegraphics[keepaspectratio]{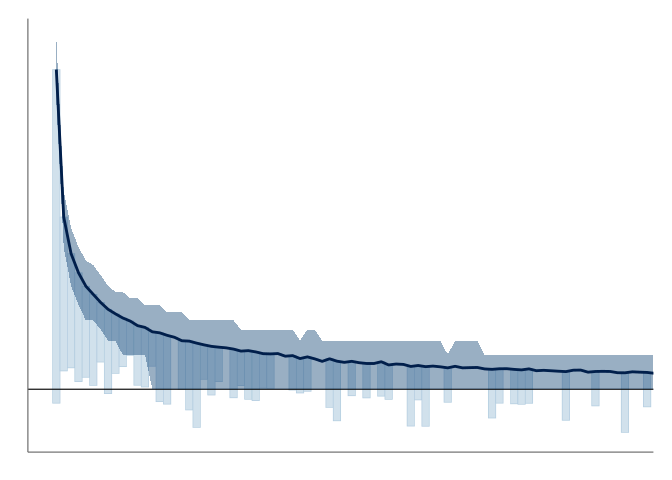}}

}

\subcaption{\label{fig-hanging-rootogram}Hanging Rootogram}

\end{minipage}%
\begin{minipage}{0.33\linewidth}

\centering{

\pandocbounded{\includegraphics[keepaspectratio]{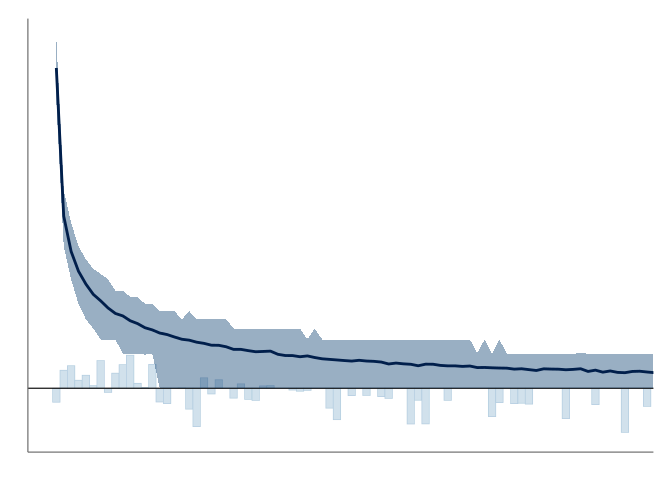}}

}

\subcaption{\label{fig-suspended-rootogram}Suspended Rootogram}

\end{minipage}%

\caption{\label{fig-rootogram-examples}The three types of rootograms
introduced by Kleiber and Zeileis
(\citeproc{ref-kleiber_visualizing_2016}{2016}), here with an additional
darker ribbon representing the predictive uncertainty. a) The standing
rootogram displays a bar graph of the observed counts overlaid with the
predictive expectations and their 90\% credible intervals. b) In the
hanging rootogram, the bar plot of the observed counts is hanging from
the predictive expectation, visualizing the average predictive error as
the empty space between the x-axis and the bars. c) In the suspended
rootogram, only the difference between the prediction and the
observation is displayed as bars.}

\end{figure}%

Kleiber and Zeileis (\citeproc{ref-kleiber_visualizing_2016}{2016})
proposed rootograms to be particularly useful in diagnosing issues, such
as over-dispersion and excess zeros, in count data models.

Traditionally, in a rootogram the frequencies of the observed counts are
visualized with a bar plot, and predictive expected frequencies as a
connected line plot. Importantly, the frequencies on the vertical axis
are plotted on a square root scale to emphasize counts with low
frequencies, occurring often in the tails of the distribution (see for
example, \citeproc{ref-tukey_graphic_1972}{Tukey 1972};
\citeproc{ref-kleiber_visualizing_2016}{Kleiber and Zeileis 2016}). As
shown in Figure~\ref{fig-standing-rootogram}, one can include the visual
representations of the predictive uncertainty without overly
complicating the visualization. Aside from the \emph{standing} rootogram
in figure Figure~\ref{fig-standing-rootogram}, two common variations are
the hanging and suspended rootogram, shown in
Figure~\ref{fig-hanging-rootogram} and
Figure~\ref{fig-suspended-rootogram} respectively. These alternatives
offer a more pattern focused approach to inspecting the possible
differences between the observed and predicted values.

In the hanging rootogram, the observed frequencies are drawn hanging
form the predictive mean, placing the lower end of the bar at the
difference of the predicted mean and the observation,
\(y_{pred} - y_{obs}\). This end is then directly readable as over- or
underestimation, depending if the bar is above or below the horizontal
axis.

The suspended rootogram instead visualizes the residual,
\(y_{obs} - y_{pred}\), as bars with their other end at the horizontal
axis of the visualization.

\begin{figure}

\begin{minipage}{0.50\linewidth}

\centering{

\pandocbounded{\includegraphics[keepaspectratio]{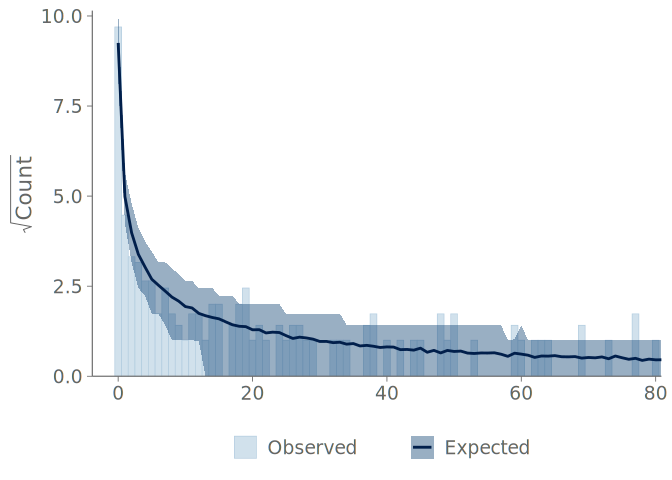}}

}

\subcaption{\label{fig-standing-rootogram-comparison}Standing Rootogram}

\end{minipage}%
\begin{minipage}{0.50\linewidth}

\centering{

\pandocbounded{\includegraphics[keepaspectratio]{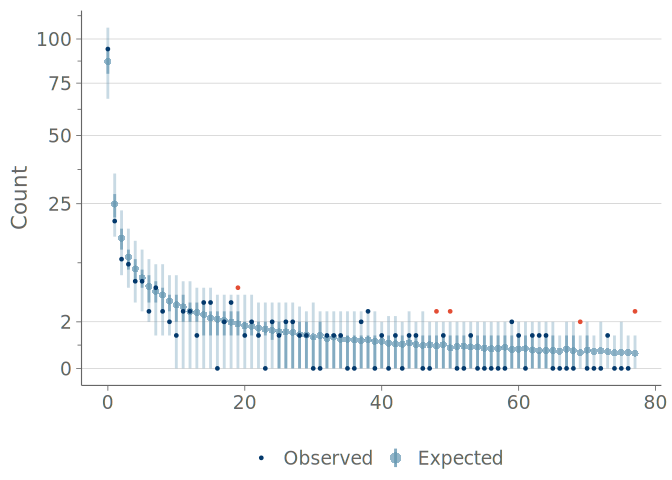}}

}

\subcaption{\label{fig-discrete-rootogram}Discrete Rootogram}

\end{minipage}%

\caption{\label{fig-new-rootogram-comparison}A side-by-side comparison
of the commonly used standing rootogram with credible intervals, and our
suggested version which emphasizes discreteness of count data. a)
Displays the square roots of the count frequencies, and visualizes the
model predictions through a continuous line and ribbon. b) Displays the
unmodified counts on the vertical axis, while using square root scaling
for the axis, thus enabling direct reading of the count frequencies. The
predictions are displayed as light blue points and interval lines. The
observations are overlaid with darker points, except the observations
outside the predictive credible intervals are highlighted in red. Again,
some points are expected to fall outside the credible intervals.}

\end{figure}%

Figure~\ref{fig-discrete-rootogram} shows our proposed modification to
the traditional rootogram. This version has two main differences to the
previously used visualizations; first, it emphasizes the discreteness of
the predictions by showing the predictive frequencies as points and
point-wise credible intervals, instead of connecting these into lines
and filled areas. For overlaying, we prefer to show the observations as
points and use color to highlight the observations falling outside the
credible intervals. Although some values are expected to fall outside
the credible intervals, these are often of interest to the practitioner
for identifying possible issues. The second important distinction is how
we implement the square root scaling on the vertical axis---instead of
the traditional method of having an axis with equally spaced tick marks
displaying the square roots of the frequencies---we transform the axis
to square root scale and display the untransformed frequencies (see
comparison in Figure~\ref{fig-new-rootogram-comparison}). We recommend
the use of axis ticks or horizontal lines as visual cues of the
non-linear scaling. This second change allows for a more natural reading
of the frequencies shown on the vertical axis of the plot.

\section{Visual predictive checks for binary
data}\label{sec-visual-predictive-checks-for-binary-data}

\begin{figure}

\begin{minipage}{0.33\linewidth}

\centering{

\pandocbounded{\includegraphics[keepaspectratio]{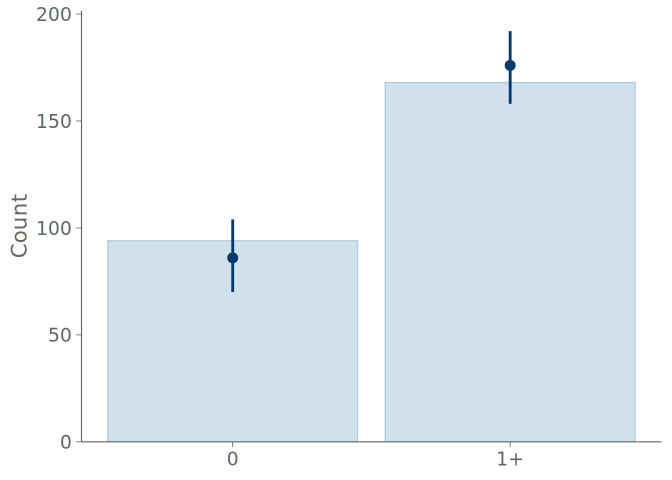}}

}

\subcaption{\label{fig-binomial-bars}Bar graph}

\end{minipage}%
\begin{minipage}{0.33\linewidth}

\centering{

\pandocbounded{\includegraphics[keepaspectratio]{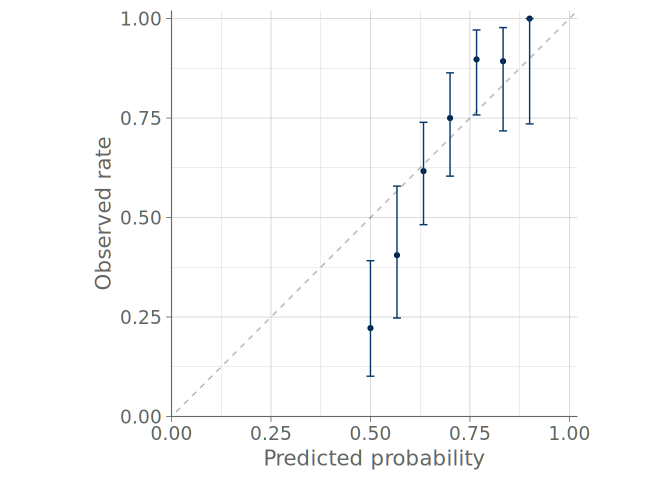}}

}

\subcaption{\label{fig-binomial-binned-calibration}Binned calibration
plot}

\end{minipage}%
\begin{minipage}{0.33\linewidth}

\centering{

\pandocbounded{\includegraphics[keepaspectratio]{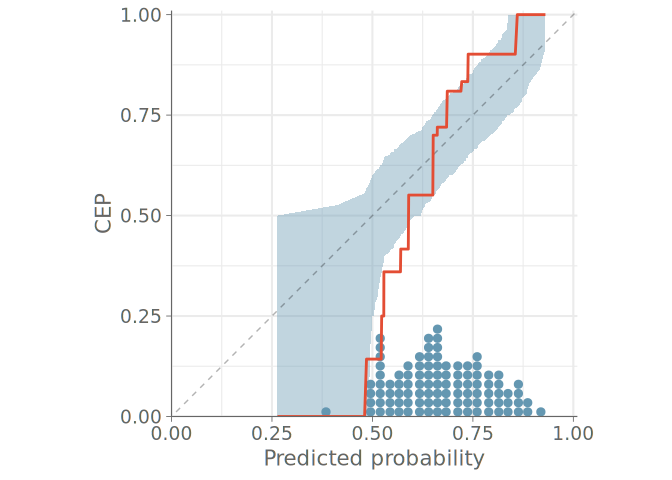}}

}

\subcaption{\label{fig-binomial-pava-calibration}PAV-adjusted
calibration plot}

\end{minipage}%

\caption{\label{fig-calibration-of-event-probability}Calibration of
estimated probability of observing a non-zero number of roaches. (a) \&
(b) commonly used bar and binned calibration plots, (c) the PAV-adjusted
calibration plot, with 95\% consistency bands computed from predictive
samples.}

\end{figure}%

Assessing the calibration of binary predictions is a common task made
difficult by the discreteness of the observed outcome. Below, we walk
through two usual approaches, and introduce an alternative offered by
PAV-adjusted calibration plots
(\citeproc{ref-dimitriadis_stable_2021}{Dimitriadis, Gneiting, and
Jordan 2021}).

We continue using the roaches example introduced in
Section~\ref{sec-count-data}, and consider three models for predicting
if one or more roaches will be observed:

\begin{enumerate}
\def\labelenumi{\arabic{enumi}.}
\tightlist
\item
  an intercept-only model estimating the overall probability of
  observing one or more roaches,
\item
  the negative binomial regression model from
  Section~\ref{sec-count-data}.
\item
  a zero-inflated version of Model 2 using the covariates to
  additionally model the probability of not observing any roaches.
\end{enumerate}

\subsection{Bar graphs}\label{bar-graphs}

A bar graph consists of bars representing the frequencies of the
observed values, and point intervals of the predictive means and central
quantiles. These elements are positioned over each state of the
observation distribution. Bar graphs are a common choice when
summarizing discrete binary, categorical, or ordinal data and
predictions. Although bar graphs can work well for predictive checking
as part of rootgrams for count data, they are less useful when there are
a small number of discrete values. In the extreme case of binary data,
even a model with one parameter corresponding to the proportion of one
class, can perfectly model the proportion, and the overlaid posterior
predictive checking bar plots are useless (and KDE plots are even worse
with misleading smoothing). Additionally, as shown in
Figure~\ref{fig-ppc-bars} and Figure~\ref{fig-binned-calibration-plots}
below, any assessment that is more advanced than inspecting the relative
frequencies of the discrete states, is impossible with the bar graphs.
For example, under- and overconfidence, and symmetric confusion between
states, is not visible in a graph.

Figure~\ref{fig-binomial-bars} shows bar graphs summarizing the
predictions of the intercept-only and negative binomial models for
predicting the binary outcome of observing one or more roaches. Both bar
graphs show the model predictions agreeing with the observations, and no
clear difference in the predictions of the models can be seen.

\begin{figure}

\begin{minipage}{0.50\linewidth}

\includegraphics[width=0.6\linewidth,height=\textheight,keepaspectratio]{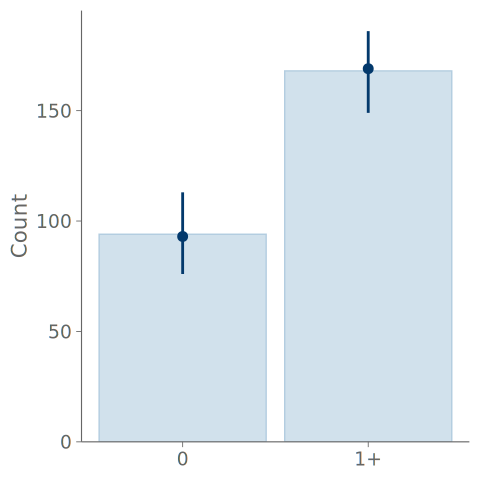}

\subcaption{\label{}Intercept-only model}
\end{minipage}%
\begin{minipage}{0.50\linewidth}

\includegraphics[width=0.6\linewidth,height=\textheight,keepaspectratio]{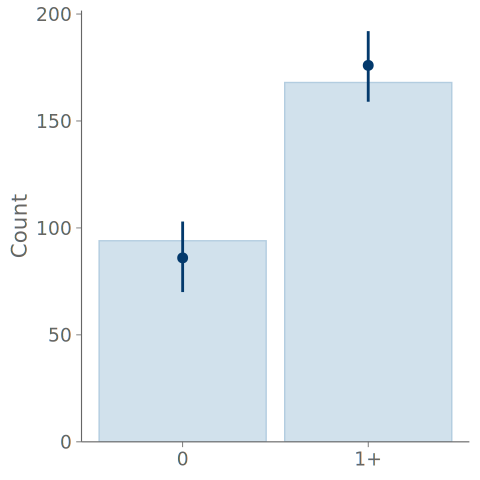}

\subcaption{\label{}Negative binomial model}
\end{minipage}%

\caption{\label{fig-ppc-bars}We cannot see a difference in the
predictive performance of the two models from the bar graphs. For both
models, the observed event frequency falls within the 95\% credible
interval with the mean predicted frequency of the intercept only model
aligning exactly with the observation. Bar plots are almost always
useless for predictive checking in case of binary data.}

\end{figure}%

\subsection{Binned calibration plots}\label{binned-calibration-plots}

What could be argued to be the first step towards assessing calibration
of the predictive probabilities in binary prediction tasks is the use of
binned calibration plots
(\citeproc{ref-niculescu-mizil_predicting_2005}{Niculescu-Mizil and
Caruana 2005}), also called calibration curves or reliability diagrams
(\citeproc{ref-degroot_comparison_1983}{DeGroot and Fienberg 1983}). In
these visualizations, the binary observations are divided into a
predetermined number of uniform bins, based on the event probabilities
predicted by the model. Comparing the mean predicted probability to the
observed event rate within each bin allows assessing the calibration, or
reliability, of the model. Typically, binned calibration plots include a
binomial confidence intervals of the observed event rates. In practice,
binned calibration plots suffer from similar drawbacks as histograms;
the choice of binning is up to the user, and can cause the resulting
plots to look drastically different, as shown by
(\citeproc{ref-dimitriadis_stable_2021}{Dimitriadis, Gneiting, and
Jordan 2021}).

Figure~\ref{fig-binomial-binned-calibration} shows the binned
calibration plots for the predictive distributions visualized previously
through bar graphs. It is immediately clear that even though the
probabilities predicted by the intercept-only model are calibrated, the
model has no ability to discriminate between the cases where one is more
or less likely to observe roaches. We can also see, that although the
negative binomial model is better at discriminating the cases, it
suffers from calibration issues and is especially under-confident at
predicting cases where no roaches will be observed.

\begin{figure}

\begin{minipage}{0.50\linewidth}

\pandocbounded{\includegraphics[keepaspectratio]{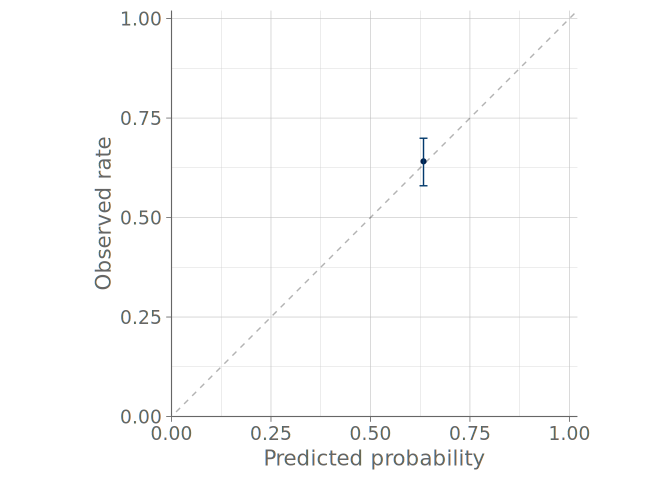}}

\subcaption{\label{}Intercept-only model}
\end{minipage}%
\begin{minipage}{0.50\linewidth}

\pandocbounded{\includegraphics[keepaspectratio]{images/03_01_binned_calibration_plot-1.png}}

\subcaption{\label{}Negative binomial model}
\end{minipage}%

\caption{\label{fig-binned-calibration-plots}The intercept only model
has no ability to discriminate between the observations, while the
negative binomial model can recognize cases where we are more likely to
observe roaches. Still, the negative binomial model the model is
under-confident in predicting cases of no roaches observed.}

\end{figure}%

\subsection{PAV-adjusted calibration
plots}\label{pav-adjusted-calibration-plots}

The ad-hoc binning decision has the potential to play a major role in
the conclusions drawn from binned calibration plots. To address this,
Dimitriadis, Gneiting, and Jordan
(\citeproc{ref-dimitriadis_stable_2021}{2021}) propose a method where
the observed binary events are replaced with conditional event
probabilities (CEP), that is the probability that a certain event occurs
given that the classifier has assigned a specific predicted probability.
To compute the CEPs, the authors use the pool adjacent violators (PAV)
algorithm (\citeproc{ref-ayer_empirical_1955}{Ayer et al. 1955}), which
provides a non-parametric maximum likelihood solution to the problem of
assigning CEPs that are monotonic with respect to the model predictions.
This monotonicity assumption is reasonable for calibrated models, where
higher predicted probabilities should correspond to higher actual event
probabilities.

To aid in the calibration assessment, we employ the point-wise
consistency bands introduced by
(\citeproc{ref-dimitriadis_stable_2021}{Dimitriadis, Gneiting, and
Jordan 2021}). These intervals show, how much variation should be
expected from the calibration plot even if the model was perfectly
calibrated.

Figure~\ref{fig-binomial-pava-calibration} shows this \emph{PAV-adjusted
calibration plot} with 95\% point-wise consistency bands computed from
the posterior predictive distribution. First, we compute the CEPs using
the posterior predictive means for each observation. Then we form the
consistency bands by using the corresponding posterior predictive draws
to determine the 95\% credible intervals for CEPs of draws from the
predictive distribution.

\begin{figure}

\begin{minipage}{0.50\linewidth}

\pandocbounded{\includegraphics[keepaspectratio]{images/03_01_pava_calibration-1.png}}

\subcaption{\label{}Negative binomial model}
\end{minipage}%
\begin{minipage}{0.50\linewidth}

\pandocbounded{\includegraphics[keepaspectratio]{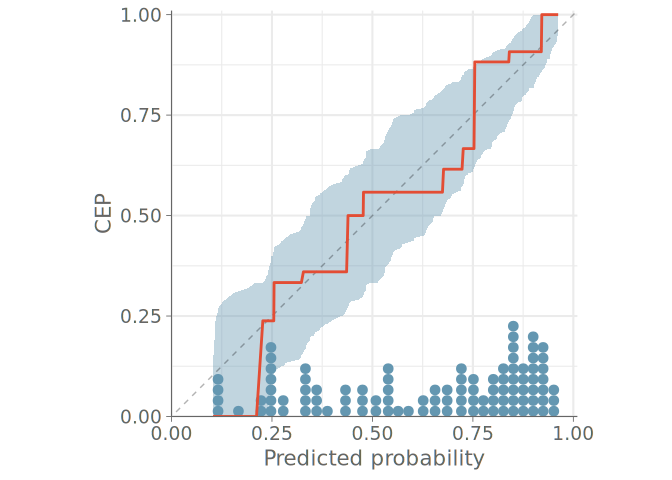}}

\subcaption{\label{}Zero-inflated model}
\end{minipage}%

\caption{\label{fig-pava-calibration}Comparison of two models for
predicting if roaches will be observed. As noted before, the negative
binomial model is under confident when predicting cases where no roaches
are observed. In the PAV-adjusted calibration plot of the zero-inflated
negative binomial model, we see the calibration curve staying within the
consistency bands, and also observe a wider range of predicted event
probabilities. Together these two qualities indicate the zero-inflated
model to be better calibrated and additionally have higher ability to
discriminate the observations with no roaches.}

\end{figure}%

\subsection{Residual plots}\label{residual-plots}

While binned and PAV-adjusted calibration plots are useful for assessing
the overall predictive calibration of a model, the behavior conditional
on individual predictors is often of interest to practitioners.

Residual plots are commonly deployed for assessing the deviation between
observations and predictions
(\citeproc{ref-gelman_regression_2020}{Gelman, Hill, and Vehtari 2020};
\citeproc{ref-gelman_bayesian_2013}{Gelman et al. 2013};
\citeproc{ref-agresti_categorical_2013}{Agresti 2013}). As with binned
calibration plots, binning is used for transforming discrete outcomes
into relative event frequencies.

\begin{figure}

\centering{

\includegraphics[width=0.9\linewidth,height=\textheight,keepaspectratio]{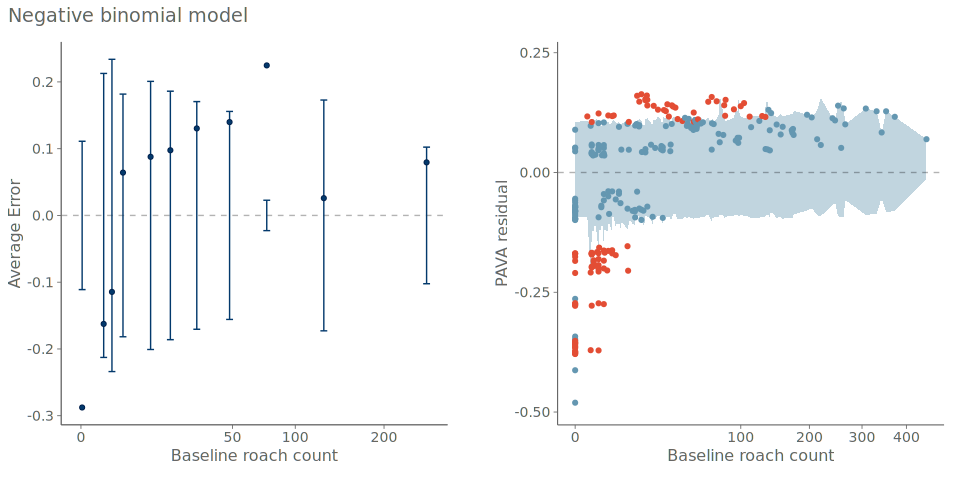}

\includegraphics[width=0.9\linewidth,height=\textheight,keepaspectratio]{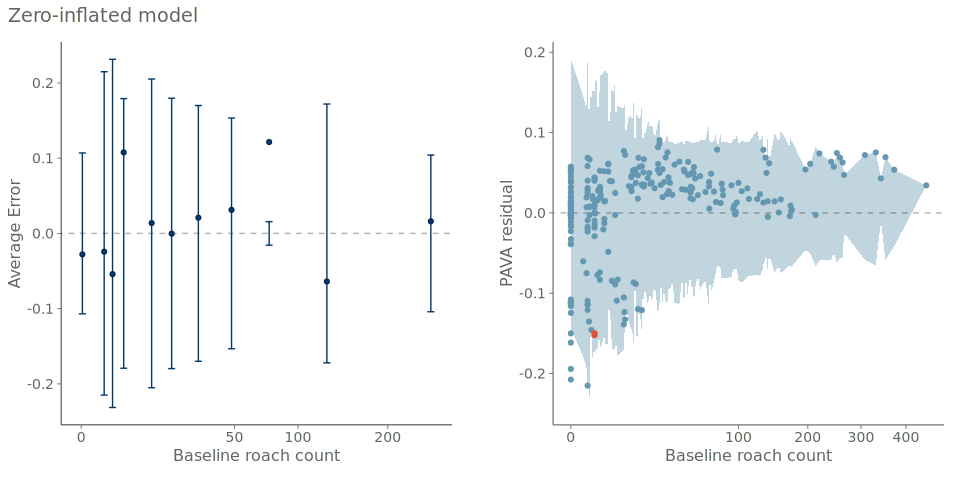}

}

\caption{\label{fig-residuals}Residual plots of the two models against
the baseline roach count observed at the start of the study. The binned
calibration plots on the left are commonly used, while the PAV-adjusted
residuals on the right offer a non-parametric alternative that doesn't
require a binning choice. Here, we use the consistency bands of the
PAV-adjusted calibration plot in Figure~\ref{fig-pava-calibration} and
highlight the observations falling outside these intervals. From both
visualizations, one can see the tendency of the simpler model without
zero-inflation to overestimate the probability of observing roaches,
when the baseline roach count is very low, as well as the tendency of
overestimating the probability of observing roaches after the treatment
in low to medium baseline counts.}

\end{figure}

Again, the ad-hoc choice of binning may result in vastly different
plots, and thus we propose the use of PAV-adjusted CEPs, computed with
regards to the predicted probabilities, in place of the discrete
observations. Figure~\ref{fig-residuals} shows an example of a
PAV-adjusted residual plot for the roaches example.

\section{Visual predictive checks for categorical
data}\label{sec-visual-predictive-checks-for-categorical-data}

When the number of categories is relatively low, the methods presented
for the binary case in
Section~\ref{sec-visual-predictive-checks-for-binary-data} extend to the
categorical case. In the same way as for binary data, even the simplest
model is likely to have such parameters that the predictive
probabilities match the observed frequencies and bar plots are useless.
However, we can use the calibration of the binomial one-versus-others
(OVO) distributions for each of the categorical cases. With a large
number of categories, investigating one plot for each OVO calibration
can be unwieldy, but scalar summaries of miscalibration for each
category could be used to filter the set of individual cases before
graphical inspection.

\begin{figure}

\centering{

\includegraphics[width=0.6\linewidth,height=\textheight,keepaspectratio]{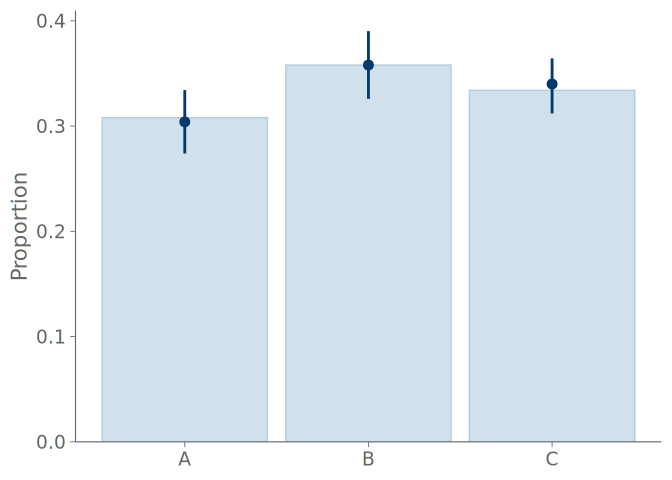}

}

\caption{\label{fig-categorical-bar-graphs}A bar graph of the observed
counts overlaid with posterior predictive means and 95\% credible
intervals. As in the earlier examples of using bar charts, everything
seems to be in order with the model predictions.}

\end{figure}

\begin{figure}

\begin{minipage}{0.50\linewidth}
\pandocbounded{\includegraphics[keepaspectratio]{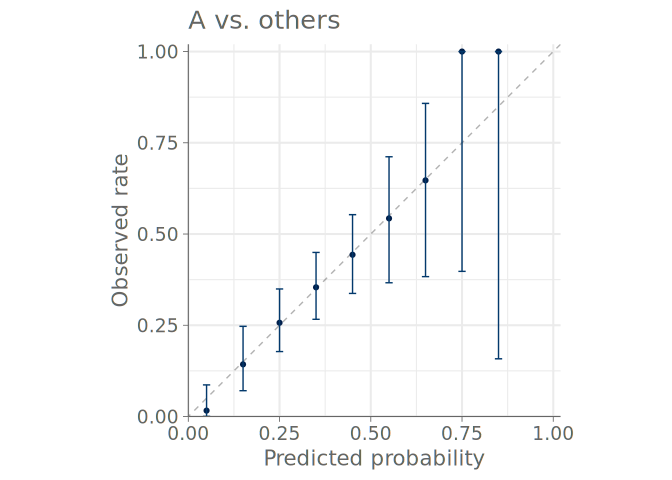}}\end{minipage}%
\begin{minipage}{0.50\linewidth}
\pandocbounded{\includegraphics[keepaspectratio]{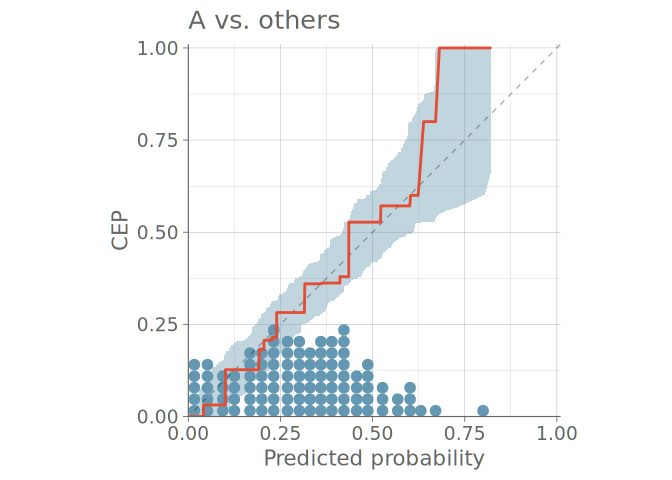}}\end{minipage}%
\newline
\begin{minipage}{0.50\linewidth}
\pandocbounded{\includegraphics[keepaspectratio]{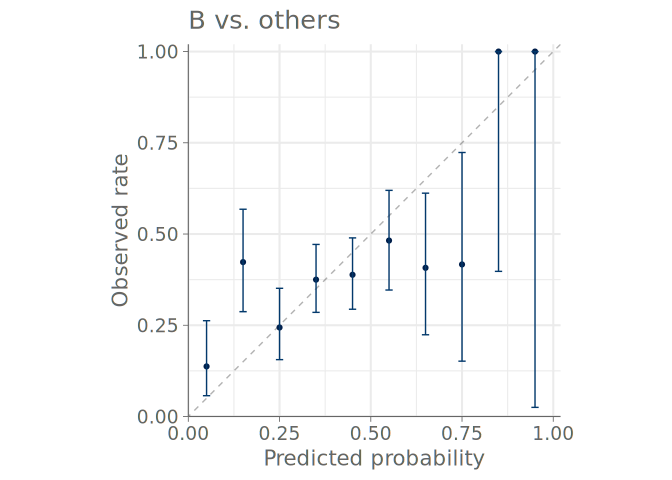}}\end{minipage}%
\begin{minipage}{0.50\linewidth}
\pandocbounded{\includegraphics[keepaspectratio]{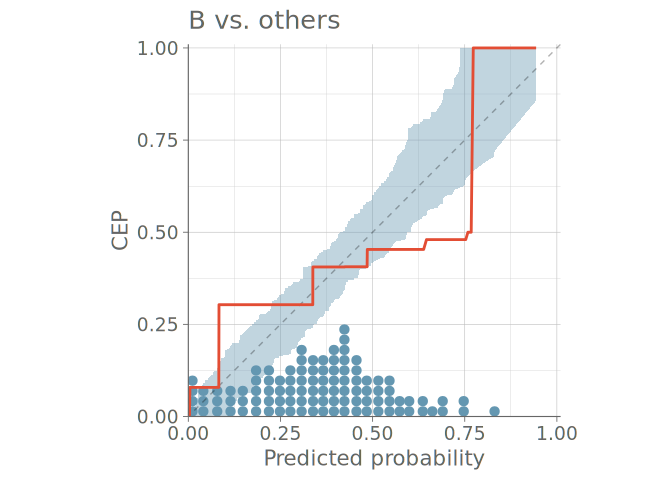}}\end{minipage}%
\newline
\begin{minipage}{0.50\linewidth}
\pandocbounded{\includegraphics[keepaspectratio]{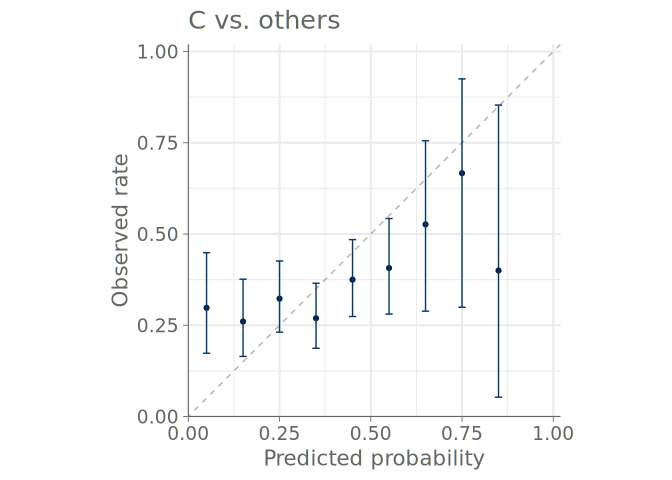}}\end{minipage}%
\begin{minipage}{0.50\linewidth}
\pandocbounded{\includegraphics[keepaspectratio]{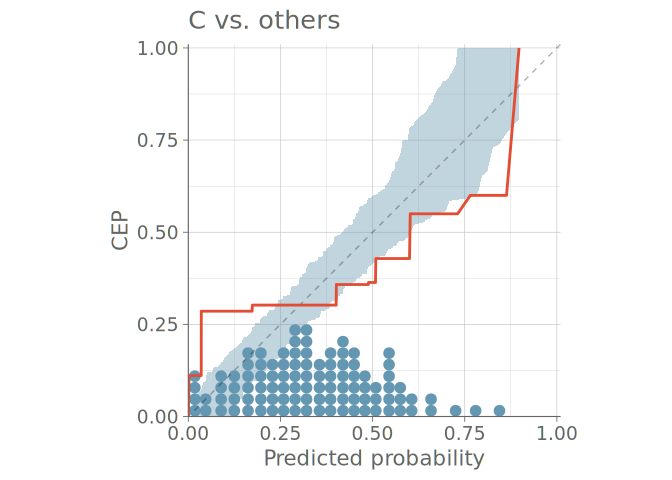}}\end{minipage}%

\caption{\label{fig-categorical-ovo-reliability-diagrams}Comparison
between binned calibration plots and PAV-adjusted calibration plots for
one versus others comparisons. We see that the predictions are well
calibrated for class A, but the model seems to be confusing observations
of classes B and C. The sharp rises at the extreme predictions indicate
that the model is over-confident in its predicted probabilities for
these classes.}

\end{figure}

Figure~\ref{fig-categorical-bar-graphs} shows a predictive check using
bar graphs for a predictive model with three distinct categories.
Figure~\ref{fig-categorical-ovo-reliability-diagrams} shows both binned
and PAV-adjusted calibration plots for the same model. No immediate
calibration issues are visible for the model when predicting category A,
but one can also see that cases, where the model would estimate the
probability of any single category to be larger than \(0.7\) are rare --
only one to three quantile dots are at values larger than \(0.7\).
Additionally, both plots types show that the model confuses classes B
and C, and is over confident in its predictions concerning these
classes.

\section{Visual predictive checks for ordinal
data}\label{sec-visual-predictive-checks-for-ordinal-data}

Models for ordinal data often have such parameters that even the
simplest model predictive probabilities match the observed frequencies
and bar plots are useless (KDE plots are even worse with misleading
smoothing). In the case of ordinal predictions, instead of looking at
the OVO distributions, we can assess the calibration of the cumulative
conditional event probabilities. This allows us to again use the toolset
presented for binomial predictive checks in
Section~\ref{sec-visual-predictive-checks-for-binary-data}, now
resulting in \(M-1\) plots, where \(M\) is the number of possible
distinct discrete cases.

\begin{figure}

\centering{

\includegraphics[width=0.6\linewidth,height=\textheight,keepaspectratio]{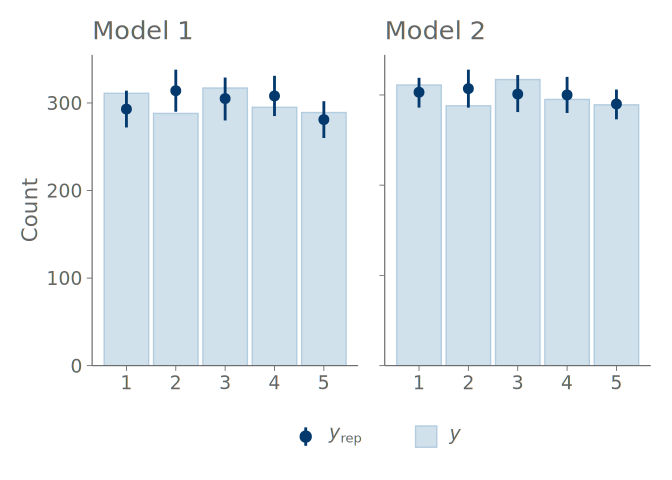}

}

\caption{\label{fig-ordinal-bar-graphs}Again, the bar graph doesn't
indicate meaningful difference between the predictive distributions.}

\end{figure}

\begin{figure}

\begin{minipage}{0.50\linewidth}
\pandocbounded{\includegraphics[keepaspectratio]{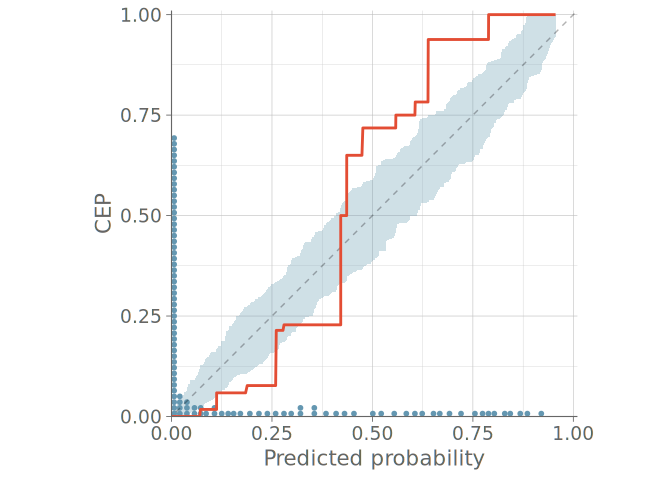}}\end{minipage}%
\begin{minipage}{0.50\linewidth}
\pandocbounded{\includegraphics[keepaspectratio]{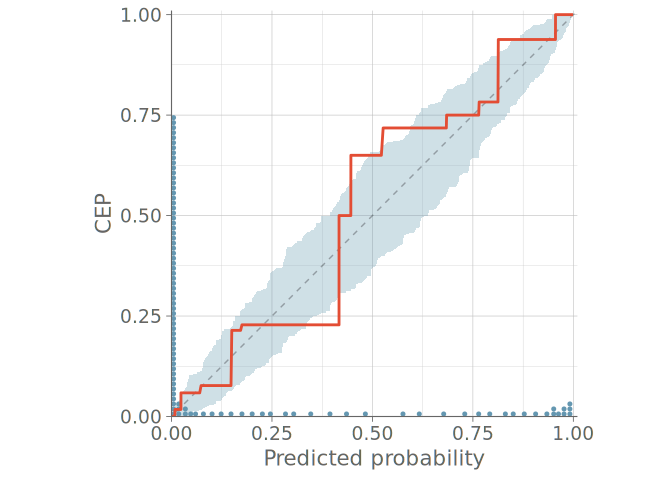}}\end{minipage}%
\newline
\begin{minipage}{0.50\linewidth}
\pandocbounded{\includegraphics[keepaspectratio]{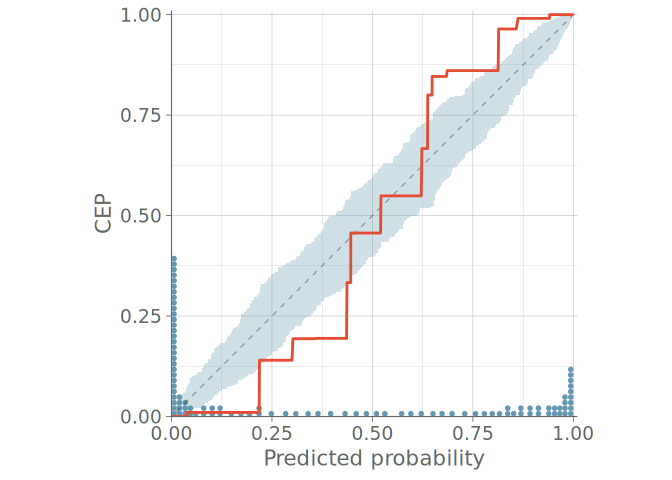}}\end{minipage}%
\begin{minipage}{0.50\linewidth}
\pandocbounded{\includegraphics[keepaspectratio]{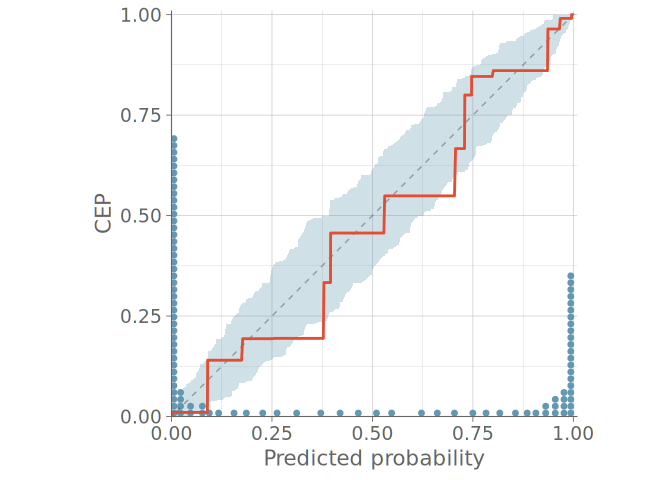}}\end{minipage}%
\newline
\begin{minipage}{0.50\linewidth}
\pandocbounded{\includegraphics[keepaspectratio]{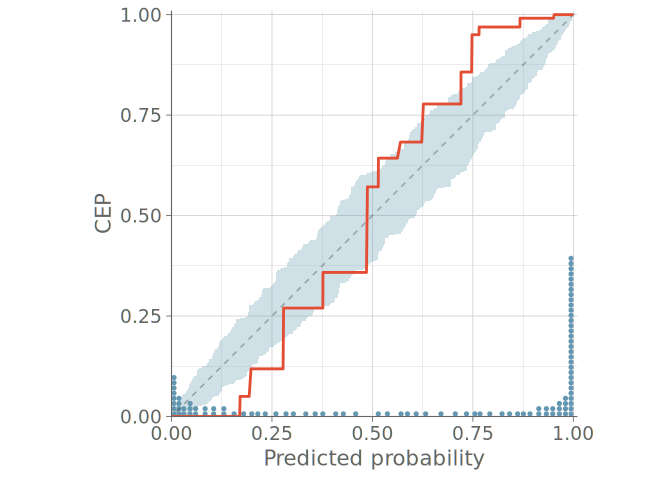}}\end{minipage}%
\begin{minipage}{0.50\linewidth}
\pandocbounded{\includegraphics[keepaspectratio]{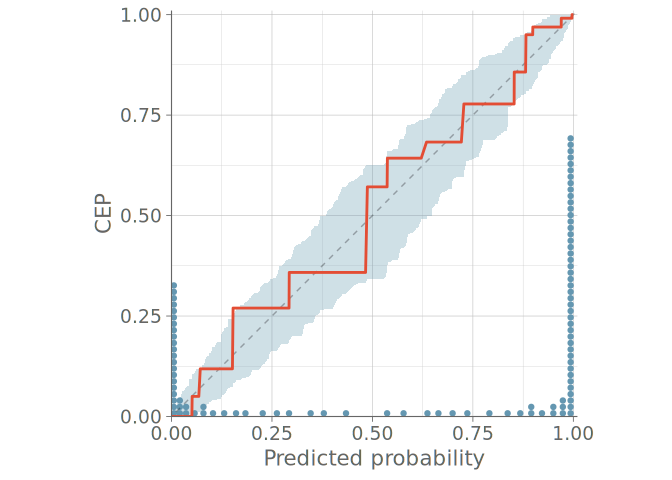}}\end{minipage}%
\newline
\begin{minipage}{0.50\linewidth}
\pandocbounded{\includegraphics[keepaspectratio]{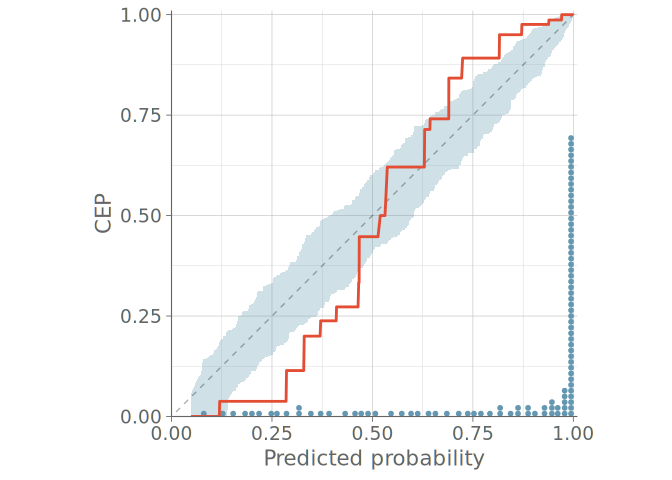}}\end{minipage}%
\begin{minipage}{0.50\linewidth}
\pandocbounded{\includegraphics[keepaspectratio]{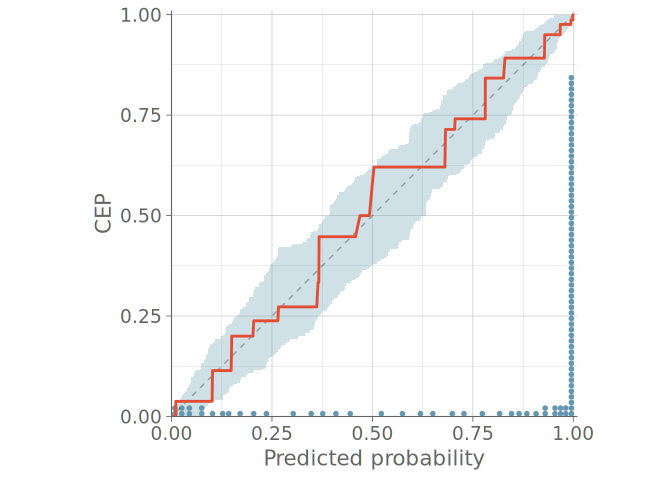}}\end{minipage}%

\caption{\label{fig-ordinal-reliability-diagrams}The side-by-side
comparison of the PAV-adjusted calibration plots for the cumulative
conditional event probabilities reveal that Model 1, containing an
implementation error, is under confident in its predictions.}

\end{figure}

Figure~\ref{fig-ordinal-bar-graphs} shows a PPC using bar graphs for two
models without any sign of calibration issues in either predictive
distribution. Figure~\ref{fig-ordinal-reliability-diagrams} shows the
PAV-adjusted calibration plots for each cumulative event probability.
The diagrams for model 1 on the left show a calibration issue, with the
calibration curves exhibiting an S-shape, indicative of under confidence
in predicted probabilities.

\begin{figure}

\centering{

\includegraphics[width=0.6\linewidth,height=\textheight,keepaspectratio]{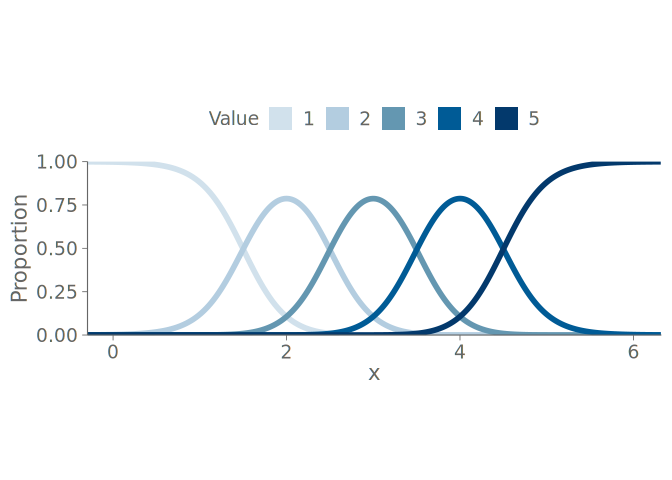}

}

\caption{\label{fig-ordinal-data-densities}True proportions of the five
possible values given value of the covariate \(x\).}

\end{figure}%

\section{Discussion}\label{discussion}

In this paper we discussed the current options for graphical predictive
checking in Bayesian model checking workflows. As discussed, the type of
data being modelled, and the type of model being applied are key
considerations when choosing a particular approach. We strongly
recommend that practitioners carefully consider the underlying
assumptions being made by a chosen visualization and perform some form
of diagnostic-check, such as the discussed PIT-ECDF, if unsure.

To summaries our recommendations, KDE plots are primarily only
recommended for when a modeller is confident that the distribution is
smooth (no point-masses or steps), or if there are many unique discrete
values. If a KDE plot is going to be used for a bounded distribution,
boundary corrections methods should be used, and we still recommend
goodness-of-fit checking. Failing this, if the distribution has a
continuous component, we recommend a quantile dot plot. For count
distributions, we recommend a modified rootogram, plotting points and
credible intervals instead of smoothed limits. For unordered discrete
data, we recommend using a PAV-adjusted One-Versus-Others calibration
plot. Finally, for ordered discrete data, we also recommend a
PAV-adjusted calibration, but evaluating the cumulative conditional
event probabilities.

It must be emphasized that these are only general recommendations. As we
have presented above, these recommended plots may still fail to identify
miscalibration or other goodness-of-fit issues in some situations.
Careful consideration of model assumptions and application of
diagnostics should always form part of the modeller's standard practice.
Many of our recommendations can be taken into account in the visual
predictive checking software, for example, by giving a warning if a user
tries to use overlaid KDE plot or bar plot for checking a model in case
of binary data.

\section*{Acknowledgements}\label{acknowledgements}
\addcontentsline{toc}{section}{Acknowledgements}

We acknowledge the support by the Research Council of Finland Flagship
programme: Finnish Center for Artificial Intelligence and Research
Council of Finland project (340721).

We would also like to thank the maintainers and contributors of the Open
Source software used in creating the visualizations and examples
presented in this article: R Core Team (\citeproc{ref-package_r}{2021});
Wickham (\citeproc{ref-wickham_ggplot2_2016}{2016}), Kay
(\citeproc{ref-kay_ggdist_2023}{2023}), and Gabry and Mahr
(\citeproc{ref-package_bayesplot}{2024b}) for their excellent
visualization packages; Narasimhan et al.
(\citeproc{ref-package_cubature}{2024}) for implementation of the
numerical integration used in the PIT computations for KDE densities;
Turner (\citeproc{ref-package_iso}{2023}), Kuhn and Max
(\citeproc{ref-package_caret}{2008}), and Dimitriadis, Gneiting, and
Jordan (\citeproc{ref-dimitriadis_stable_2021}{2021}) for implementing
functions used in making calibration plots; Bürkner
(\citeproc{ref-package_brms1}{2017}), Gabry et al.
(\citeproc{ref-package_cmdstanr}{2024}), Goodrich et al.
(\citeproc{ref-package_rstanarm}{2024}), and Stan Development Team
(\citeproc{ref-standev2018stancore}{2018}) for developing the
probabilistic programming software used in this article.

\phantomsection\label{refs}

\end{document}